\documentclass[sigconf]{acmart}
\usepackage{graphicx} % Required for inserting images
\usepackage{multirow}
\usepackage{tabularray}
\usepackage{multicol}
\usepackage{booktabs}
\usepackage{appendix}
\usepackage{caption}
\usepackage{subcaption}
\usepackage{enumitem}
\usepackage{array}
\usepackage{multicol} % in preamble
\usepackage{longtable}   % in preamble
\usepackage{booktabs} 

\usepackage{listings}
\definecolor{blue}{HTML}{2A6EFF} % blue

\setcopyright{none}
\copyrightyear{YYYY}
\acmYear{YYYY}
\acmDOI{XXXXXXX.XXXXXXX}
\acmConference[Conference acronym 'XX]{In Submission to CHI 2026}{June 03--05, 2018}{Woodstock, NY}

\title[Prototyping Digital Social Spaces through Metaphor-Driven Design]{Prototyping Digital Social Spaces through Metaphor-Driven Design: Translating Spatial Concepts into an Interactive Social Simulation}
\author{Yoojin Hong}
\email{dbwk18@kaist.ac.kr}
\affiliation{
  \institution{KAIST}
  \streetaddress{291 Daehak-ro, Yuseong-gu}
  \city{Daejeon}
  \country{South Korea}
  \postcode{34141}
}

\author{Martina Di Paola}
\email{martinadipaola@kaist.ac.kr}
\affiliation{
  \institution{KAIST}
  \streetaddress{291 Daehak-ro, Yuseong-gu}
  \city{Daejeon}
  \country{South Korea}
  \postcode{34141}
}

\author{Braahmi Padmakumar}
\email{pjbrahmi@kaist.ac.kr}
\affiliation{
  \institution{KAIST}
  \streetaddress{291 Daehak-ro, Yuseong-gu}
  \city{Daejeon}
  \country{South Korea}
  \postcode{34141}
}

\author{Hwi Joon Lee}
\email{lee.hw@northeastern.edu}
\affiliation{
  \institution{Northeastern University}
  \streetaddress{360 Huntington Avenue}
  \city{Boston}
  \country{United States of America}
  \postcode{02115}
}

\author{Mahnoor Shafiq}
\email{mahnoorshafi13@kaist.ac.kr}
\affiliation{
  \institution{KAIST}
  \streetaddress{291 Daehak-ro, Yuseong-gu}
  \city{Daejeon}
  \country{South Korea}
  \postcode{34141}
}

\author{Joseph Seering}
\email{seering@kaist.ac.kr}
\affiliation{
  \institution{KAIST}
  \streetaddress{291 Daehak-ro, Yuseong-gu}
  \city{Daejeon}
  \country{South Korea}
  \postcode{34141}
}

\date{September 2025}

\begin{document}

\begin{abstract}
Social media platforms are central to communication, yet their designs remain narrowly focused on engagement and scale. While researchers have proposed alternative visions for online spaces, these ideas are difficult to prototype within platform constraints. In this paper, we introduce a metaphor-driven system to help users imagine and explore new social media environments. The system translates users' metaphors into structured sets of platform features and generates interactive simulations populated with LLM-driven agents. To evaluate this approach, we conducted a study where participants created and interacted with simulated social media spaces. Our findings show that metaphors allow users to express distinct social expectations, and that perceived authenticity of the simulation depended on how well it captured dynamics like intimacy, participation, and temporal engagement. We conclude by discussing how metaphor-driven simulation can be a powerful design tool for prototyping alternative social architectures and expanding the design space for future social platforms.
\end{abstract}




\begin{teaserfigure}
    \centering
    \includegraphics[width=\textwidth]{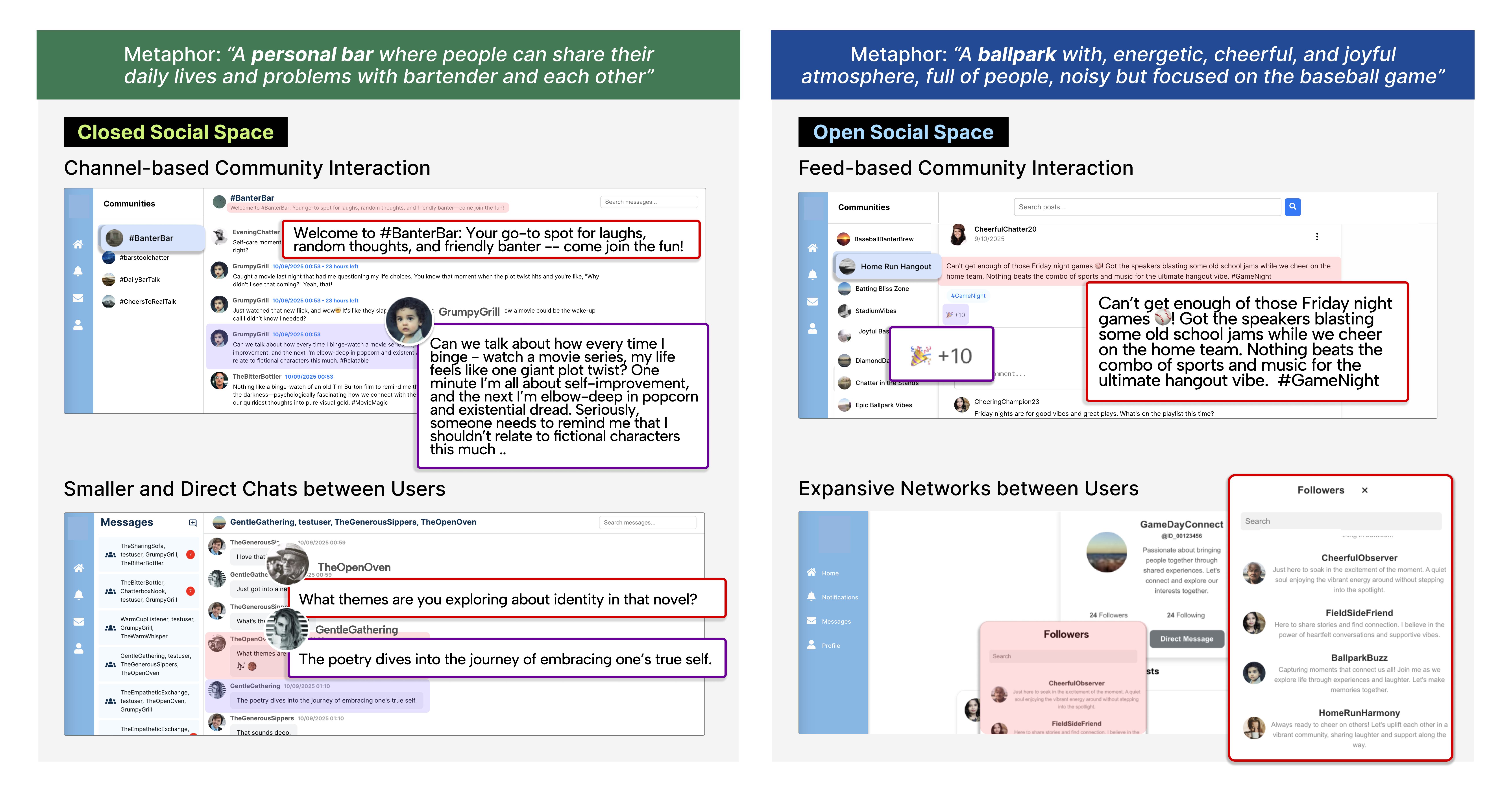}
    \caption{Example simulated digital social spaces generated from two participant-provided metaphors: ``A personal bar where people share their daily lives'' and ``A ballpark with an energetic, cheerful, and joyful atmosphere''. The left panel shows a more intimate configuration with distinct channels, smaller and direct chats, and conversational agents that mirror the kinds of interactions expected in closed settings. The right panel illustrates a more expansive layout with multiple feeds, broader community connections, and visible networks that reflect the openness and scale of the imagined ballpark setting.}
    \Description{The image shows examples of the system interface for the simulated social spaces. On the left in a green bar at the top is the first metaphor: "A personal bar where people can share their daily lives and problems with bartender and each other". Below that shows the system as a closed social space, with the top image showing a channel-based community interaction among the simulated agents. The channel is called #BanterBar, a go-to spot for laughs, random thoughts, and friendly banter. The agents are talking about a movie they have seen. Below is an example of smaller and direct chats between users, showing two agents talking about a poetry book. On the right side in a blue bar at the top is the second metaphor: "A ballpark with energetic, cheerful, and joyful atmosphere, full of pepole, noisy but focused on the baseball game. Below that shows the system as an open social space, with the top image showing a feed-based community interaction among the simulated agents. There is a feed with a user making a post about baseball game night. Below is an example of expansive networks between users, showing many followers for a baseball game community.}
    \label{fig:system-figure}
\end{teaserfigure} 

% \begin{teaserfigure}
%     \centering
%     \includegraphics[width=\textwidth]{figures/system-study-overview.png}
%     \caption{Overview of system pipeline, user study process, and research questions. This figure illustrates the end-to-end workflow of our study. The System Development phase (top) begins with (1) user-provided spatial metaphors, (2S) which are expanded into structured metaphor descriptions, (3S) translated into platform features via a 3-level taxonomy, and (4S) instantiated as interactive social media simulations populated with LLM-driven agents by composing modular UI components into a simulation environment.
%     The User Study Process (bottom) evaluates each phase to address four research questions (RQ1–RQ4) through: (1) user input of a spatial metaphor, (2U) user-authored social attribute descriptions, (3U) feature alignment ratings, and (4U) post-interaction feedback.
%     These phases correspond to the folowing four research questions: RQ1 - how users conceptualize social spaces; RQ2 - alignment of social attributes; RQ3 - how well simulated features represent users' metaphors; and RQ4 - perceived authenticity of the simulated experience.}
%     \label{fig:system-study-overview}
% \end{teaserfigure} 

\maketitle

\section{Introduction}

Over the past two decades, social media platforms have become dominant infrastructures for building social connections and exchanging information. 
However, the design of these platforms has remained constrained by a relatively narrow set of conventions~--~feeds, likes, follows, and algorithmically sorted content~--~prioritizing engagement metrics over meaningful interaction~\cite{backstrom2016serving, narayanan2023understanding, scharlach2023value}.
While early imaginaries of the internet envisioned diverse and participatory virtual ``places'' for community, the actual architectures of platforms have emphasized scalability rather than healthy or rich interpersonal connections.
Despite growing interest in designing alternative systems that support more constructive and value-sensitive online environments~\cite{bernstein2023embedding}, these efforts have struggled to move beyond the dominant paradigm to articulate and implement functionally new social models. As Henri Lefebvre famously argued, \textit{``Space is social: it involves assigning more or less appropriated places to social relations... social space has thus always been a social product''}~\cite{lefebrve1991production}. 
Yet in digital environments, the spatial dynamics that facilitate organic social interaction are often diminished. 

To envision more diverse and value-sensitive online social environments, we turn to metaphor as a generative design tool. 
Metaphors enable individuals to project familiar spatial and social experiences into new contexts, offering conceptual frameworks for imagining alternative social architectures. 
In this paper, we refer to such metaphorical framings as~\textit{spatial metaphors}. 
Prior work has shown how metaphors support the formation of mental models~\cite{lakoff2008metaphors, dourish2001where} and design ideation by grounding abstract system behaviors in familiar, experiential language~\cite{sengers2005reflective, hook2018designing}. 
In the context of social media, spatial metaphors provide users with a means to articulate their relational values and expectations, enabling the imagination of social spaces that move beyond the defaults of current platform design.

One core challenge in operationalizing metaphor-driven design lies in testing how imagined social spaces might function at scale~--~particularly how interaction dynamics would unfold among diverse users in real-time settings.
Simulations have long been used in HCI and social science to model user behavior and predict outcomes of system design~\cite{jacko2012human, gilbert2005simulation}, with agent-based modeling serving as a popular approach for leveraging computational power to simulate social phenomena~\cite{de2014agent}. 
Recent advancements in large language models (LLMs) and generative agent-based simulations (see, e.g., ~\cite{park2022simulacra, park2023generative}) have further expanded the possibilities for prototyping social systems, enabling dynamic, interactive environments where user behaviors and platform features can be rapidly instantiated and tested.
While recent work has leveraged simulations to investigate social dynamics~--~such as information diffusion~\cite{zhou2020realistic} and opinion polarization~\cite{haque2023understanding}~--~these studies primarily aim to validate theoretical assumptions or predict behavioral outcomes.
Our work positions simulation as a design-oriented tool: a sandbox environment that transforms users’ metaphor-driven imaginations into interactive prototypes, allowing them to experience and evaluate how closely the resulting interactions align with their envisioned values and social expectations.

In this paper, we present a system that supports users in imagining, generating, and interacting within new social media spaces derived from their own spatial metaphors. 
The system translates key elements of each user-provided metaphor~--~such as atmosphere, interaction patterns, and relational dynamics~--~into concrete platform features, including content flow, visibility settings, and modes of communication, generating an interactive prototype that reflects the metaphor’s underlying social logic. 
The resulting environment is populated by LLM-driven agents designed to simulate user interactions consistent with the metaphor’s intent, allowing users to explore and reflect on how well the generated space aligns with their expectations.

To explore how our system translates user-imagined spatial metaphors into interactive social media spaces, we conducted a two-phase study using contrasting social space prompts: social spaces with high vs. low levels of contextual openness, which we refer to as \textit{open} and \textit{closed} social spaces throughout the remainder of this text. Participants were required to imagine one social space of each type in order to test the flexibility of the system in generating different spaces. We explain and justify our focus on this dimension of contextual openness in Section \ref{sec5:study-design}. Each participant described a metaphor for an open space and a closed space, reviewed the system-generated feature sets, and interacted with a simulated environment populated by LLM-driven agents generated based on the submitted metaphors. Drawing on data from user-created metaphors, surveys, and system-use interviews, we assessed what kinds of social spaces users imagine, how effectively the system captured the intended social dynamics of their metaphors, which attributes aligned or conflicted in translation, and what design elements influenced the perceived coherence of the simulated experience. We structured our study to answer four main research questions:

\begin{itemize}
    \item \textbf{RQ1:} How do people conceptualize digital social spaces through metaphor? 

    \begin{itemize}
        \item[o] What types of metaphorical dimensions do users emphasize when imagining open vs closed social spaces?
    \end{itemize}

    %\item \textbf{RQ2:} How effective is the system’s metaphor conversion pipeline in translating user-imagined social spaces into platform features? %Which social attributes tend to align well, and which ones give rise to conflicts during the translation process?

    \item \textbf{RQ2:} What are the key features necessary to blueprint a user’s imagined social space into a functional social media platform?

    \item \textbf{RQ3:} How do users' experiences in the simulated environments align or misalign with their expectations for the metaphorical space? %What design elements contribute to this alignment or misalignment?
\end{itemize}

To investigate how metaphor can support a richer investigation of potential digital social spaces, we conducted a study in four phases: (1) participants input their spatial metaphor into the system; (2) while waiting for the system to generate the simulated space, participants elaborated upon their metaphor and explained it in more depth; (3) the participants evaluated how well the  feature set proposed by the system aligned with their concept; and (4), the participants interacted with the resulting simulated space and gave feedback through surveys and interviews on how well it matched their expectations. Participants went through this process twice~--~once to imagine an open space and once to imagine a more closed space.

We find that participants were easily able to conceptualize their ideal social spaces through metaphor, with consistent distinctions emerging between open and closed environments~--~open spaces were described as engaged, energetic, and spontaneous, while closed spaces were characterized by relaxed, intimate, and extended engagement. Participants identified a mix of well- and poorly-aligned features between what they had imagined and what the system generated, highlighting both potential weaknesses of the system and also limitations of the use of metaphors for this purpose. Finally, participants reflected on their experiences interacting with the simulations, showing that perceived authenticity was closely tied to how well content flow, messaging coherence, and identity visibility mirrored the social expectations embedded in the original metaphors.  
Together, these findings demonstrate that meaningful simulation of social platforms depends on extracting metaphorical structures and carefully tuning system features to preserve the relational nuances users associate with different social spaces. 

We conclude by discussing the strengths and weaknesses of this approach. We note the important limitation that this system imagines new spaces through recombination of existing social media features (e.g., likes, direct messages, followers) rather than allowing users to imagine and instantiate completely new features, and we imagine how future systems might enable more flexible, open-ended design exploration for social spaces.

\section{Related Work}
This section reviews prior research that informs our approach. We begin by examining how design constraints in mainstream social media platforms may have contributed to recurring harms. We then discuss alternative design visions and introduce spatial metaphor as a generative lens for reimagining online social interaction. Finally, we review emerging work on simulation, focusing on LLM-driven agent-based systems as tools for prototyping and evaluating new social architectures. 

\subsection{Design Constraints leading Harms in Existing Social Media Platforms}

Social media has been developed as a way for people to connect, share, and participate in public discourse across geographic and social boundaries. Over the past two decades, platforms like Facebook, Twitter, and Instagram have become integral parts of everyday life, shaping how people maintain relationships, express identity, and access information. However, as these platforms have matured, a growing number of studies and public commentaries have raised concerns about the unintended consequences of their design.

A key concern is that many of the dominant social media architectures~--~feeds, likes, follows, and algorithmic curation~--~have been optimized primarily for engagement and scalability, often at the expense of social and psychological well-being~\cite{zuboff2019surveillance, tufekci2018twitter}. Features such as infinite scroll, public-by-default settings, and reaction-based feedback systems encourage habitual use and emotional dependency, while limiting the kinds of interpersonal dynamics that are possible~\cite{andrews2015beyond}. These structural constraints have been linked to a range of potential harms, including increased anxiety, loneliness, social comparison, and mental health struggles, especially among adolescents and younger users~\cite{twenge2018depressive, orben2019association, rideout2018social}, though the precise relationship between these phenomena remains difficult to quantify.

Current platform designs may also make it difficult for users to manage social boundaries, navigate multiple social roles, or express themselves differently in different contexts. Studies have shown that flattened audience structures and the lack of situational cues create tensions around self-presentation and privacy~\cite{boyd2014its, marwick2011social}. The design of visibility and interaction defaults often favors performativity and public expression over private or context-sensitive interaction~\cite{bucher2018if, hu2014instagram}, which in turn contributes to the erosion of interpersonal trust and the amplification of social pressure.

In sum, while social media platforms offer unprecedented communicative reach, the present range of social platform designs has resulted in a set of recurring harms. These issues underscore the need for new design approaches that center relational nuance, contextual expression, and user-defined values from the outset.

\subsection{The Role of Spatial Metaphor in Rethinking Social Media}
Recognizing the structural harms embedded in current social media, HCI researchers have proposed alternative design paradigms aimed at promoting healthier, more meaningful online experiences. Much of this work seeks to reorient design goals away from engagement and virality and toward values such as well-being, mutual support, and expressive autonomy.

Participatory and speculative design practices have been central to these efforts. Researchers have employed co-design workshops, speculative fiction, and reflective prototyping methods to engage users in envisioning new modes of digital interaction. For instance, participatory design studies with teenagers and parents have explored alternative architectures that prioritize safety, creativity, and identity exploration for young users~\cite{yarosh2016locus, sobieraj2020locus}. Similarly, value-centered design strategies suggest that platforms should be evaluated not only on usability or retention but also on how well they support long-term user flourishing. Baumer et al.~\cite{baumer2015reflective} introduced reflective informatics as a strategy to support intentional and self-aware use of social technologies. These approaches highlight the limitations of conventional social media and aim to surface alternative imaginaries that are grounded in situated needs and relational goals. However, while these works surface important value tensions and design goals, many of these interventions remain limited in scope. Most proposals are articulated in terms of isolated features or interventions within existing platforms, such as tools for boundary management~\cite{marwick2011networked}, visibility controls~\cite{gillespie2018custodians}, or modified content recommendation algorithms~\cite{bucher2018if, schmidt2020formal}, rather than providing a generative framework for reimagining platform structure at a higher level.

To move beyond these constraints, we turn to \textbf{metaphor} as a generative lens for conceptualizing and prototyping alternative social platforms. In HCI and design research, metaphors have long served as tools for scaffolding mental models and framing abstract system behavior in familiar terms~\cite{lakoff2008metaphors, sengers2005reflective, blackwell2006reification}. Foundational computing metaphors~--~such as the ``desktop,'' ``windows,'' and ``forums''~--~have shaped how users interact with digital environments by grounding novel interactions in familiar conceptual schemas~\cite{carroll1982metaphor, dourish2001where}. Dourish~\cite{dourish2001where} in particular emphasized how metaphors do not merely describe system behavior but actively shape social practice and user expectation.

We focus here on the use of metaphors in spatial and social domains, exploring how spatial metaphors can structure relational expectations and interaction norms within digital environments. Theories of place in digital systems further support the importance of metaphors in structuring social interactions, arguing that digital environments can~--~and should~--~embody social dimensions traditionally associated with physical place, including atmosphere, co-presence, and contextual interaction~\cite{harrison1996re}. By treating metaphor as a starting point for design rather than a descriptive overlay, we enable users to articulate the relational values and social dynamics they wish to see instantiated in digital space. This approach supports the exploration of entirely new social architectures~--~beyond incremental changes to existing platforms~--~and expands the design space for imagining how people might want to interact, participate, and be recognized in future social media systems.

% We extend these metaphor use into the spatial and social domains, exploring how spatial metaphors  can structure relational expectations and interaction norms within digital environments. Theories of place in digital systems further support this view, arguing that digital environments can—and should—embody social dimensions traditionally associated with physical place, including atmosphere, co-presence, and contextual interaction~\cite{harrison1996re}.  By treating metaphor as a starting point for design rather than a descriptive overlay, we enable users to articulate the relational values and social dynamics they wish to see instantiated in digital space. This approach supports the exploration of entirely new social architectures—beyond incremental changes to existing platforms—and expands the design space for imagining how people might want to interact, participate, and be recognized in future social media systems.

\subsection{Simulation as a Method for Social Media Prototyping}

Despite a growing body of work proposing alternative designs for social media platforms, a persistent challenge remains: most proposals are difficult to evaluate beyond the context of conceptual critique or small-scale prototypes. Many interventions focus on modifying isolated features of existing systems, such as visibility settings or content curation algorithms, rather than envisioning the platform as a fundamentally different kind of social environment. These incremental modifications are typically constrained by the architectural logic of mainstream platforms, making it difficult to examine how significantly new forms of interaction, governance, or community formation might emerge in practice. Moreover, constructing fully functional alternatives often demands substantial engineering resources and scalability required to test multiple directions efficiently.

Simulation offers a promising alternative for addressing these limitations. Rather than requiring fully deployed systems, simulations allow researchers to instantiate and iteratively explore imagined social environments within controlled settings. With the emergence of large language models (LLMs), this approach has gained new momentum: LLM agents can exhibit open-ended, socially-plausible behaviors that support rich multi-agent interaction. Recent surveys position LLM-based simulation as a robust methodological paradigm. Sun et al.~\cite{sun2024llmsocialsims} emphasize simulation’s potential to support scalable experimentation with social structures while mitigating the ethical and logistical constraints of traditional human-subject research. Li et al.~\cite{li2024survey} provide a comprehensive synthesis of LLM-driven simulation research, highlighting a growing emphasis on modeling society-scale interactions. Their review identifies key research directions that move beyond dyadic communication or task-oriented behavior to simulate emergent collective dynamics, opinion evolution, and social norm formation.

Building on this methodological groundwork, several recent systems use LLM-based simulation to examine complex social phenomena. SocioVerse~\cite{lee2024socioverses}, for instance, aligns LLM agents with a large-scale dataset of user profiles to simulate cross-domain behaviors in political, economic, and news contexts. Other projects adopt different technical orientations while pursuing similarly ambitious goals. Y Social~\cite{rossetti2024ysocial} constructs a digital twin of existing social media platforms, enabling fine-grained analysis of user-system interactions under varying design parameters. SimSpark~\cite{fard2024simspark} offers a real-time, lightweight simulation framework for modeling spontaneous interpersonal exchanges. In addressing broader societal challenges, Concordia~\cite{puelma2024concordia} combines LLM agents with user surveys to test interventions aimed at mitigating misinformation in decentralized social networks.

While these systems establish early analytical utility for LLM-based simulation, another body of work explores the experiential and narrative aspects of simulated social interaction. Generative Agents~\cite{park2023generative} and Social Simulacra~\cite{park2022simulacra} pioneered interactive environments in which agents engage in evolving social narratives informed by memory, goals, and environmental cues. These systems shift the focus from modeling isolated behaviors to simulating routines, highlighting the role of simulation as a medium for expressing imagination and modeling human complexity.

% While these systems establish the analytical utility of LLM-based simulation, another body of work explores its experiential and narrative aspects of social interaction. Generative Agents~\cite{park2023generative} and Social Simulacra~\cite{park2022simulacra} pioneer interactive environments in which agents engage in evolving social narratives informed by memory, goals, and environmental cues. These systems shift the focus from modeling isolated behaviors to simulating situated routines and intersubjective meaning-making, highlighting the role of simulation as a medium for imaginative speculation. Rather than abstracting away human complexity, they foreground it as a creative constraint that enables exploration of alternative social logics.

Building on this foundation, our work introduces a \textit{design-oriented perspective} to LLM-based simulation. Rather than modeling existing platforms or replicating interaction patterns for analysis, we enable users to articulate and explore novel social architectures through metaphorical description. Participants begin by specifying a spatial metaphor for a hypothetical digital social space, such as ``a research lab,'' ``a rooftop party,'' or ``a street corner during a parade'',\footnote{As this study was conducted multi-lingually, and the concept of a metaphor varies across languages~\cite{Mitić2025metaphor, Deignan2003metaphor}, the examples of spatial metaphors given here may seem variably closer to or further from the ideal case depending on the reader's perspective. We argue that part of the strength of a metaphor is that it does not need to fit a precise template, so we allowed for variation in how participants phrased their spatial metaphors.} which serves as a scaffold for defining social conditions, atmospheres, and expectations for participation. These metaphorical inputs are translated into structured platform features and instantiated with generative agents, allowing users to directly experience a version of their imagined environments. By combining metaphor-based specification with generative multi-agent simulation, we aim to expand the design space for social media systems, enabling users to envision, instantiate, and explore new social forms that might otherwise remain abstract or infeasible to test through conventional prototyping methods.
\section{System Design}\label{sec3:metaphor-pipeline}

\begin{figure*}
    \centering
    \includegraphics[width=\textwidth]{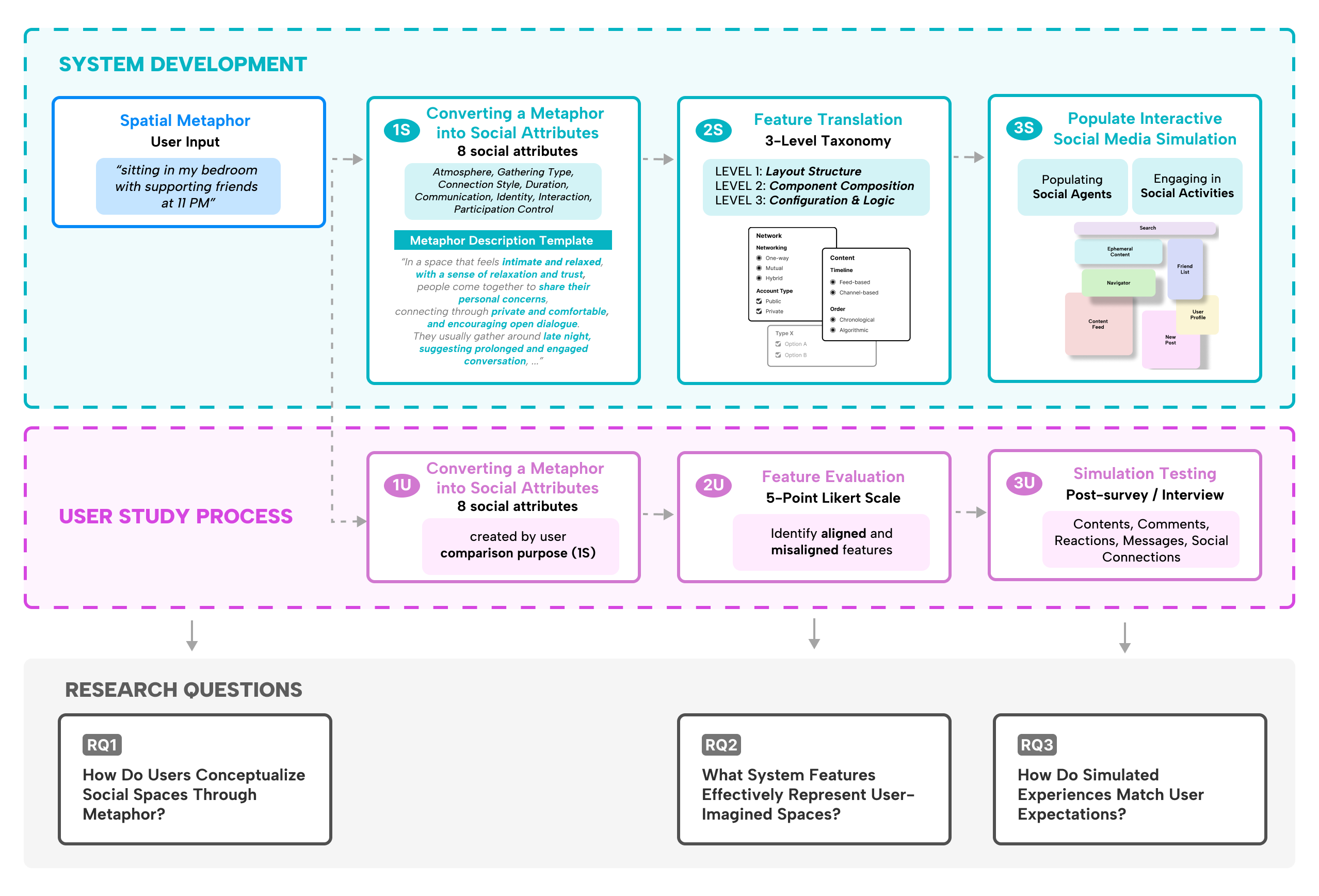}
    \caption{Overview of system pipeline, user study process, and research questions. This figure illustrates the end-to-end workflow of our study. The System Development phase (top) begins with user-provided spatial metaphors, which are expanded into structured metaphor descriptions (1S), translated into platform features via a 3-level taxonomy (2S), and instantiated as interactive social media simulations populated with LLM-driven agents by composing modular UI components into a simulation environment (3S).
    The User Study Process (bottom) evaluates each phase to address three research questions through: user-authored social attribute descriptions of the metaphor(1U), feature alignment ratings (2U), and post-interaction feedback (3U).}
    \Description{The image shows an overview of the system pipeline, user study process, and the research questions. At the top is the System Development phase, which begins with a blue box showing an example metaphor input of "sitting in my bedroom with supporting friends at 11 PM" by the user. A gray arrow pointing to the right connects to a teal box showing the metaphor converted into 8 social attributes. A gray arrow pointing to the right connects to a teal box showing the feature translation based on a three-level taxonomy. A gray arrow pointing to the right connects to a teal box that shows that the social space is populated with social agents, who engage in social activities. In the middle below is the user study process in pink. For the first pink box, the user study process shows the user creating a metaphor into social attributes. Then a gray arrow pointing to the right connects to a box showing the feature evaluation with five-point likert scale, which then connects to a box showing simulation testing being done through post-survey and interviews. Below at the bottom are the research questions in gray, aligning with the system development and user study processes in order.}
    \label{fig:system-study-overview}
\end{figure*}

In this section, we show how a metaphor can provide a structure for translating a user-imagined social space into an interactive, simulated social media environment. Here, we use the term, \textit{spatial metaphor}, to mean a physical or environmental setting that encapsulates a desired form of social interaction for a hypothetical social media space. These metaphors act as conceptual anchors that express the atmosphere, structure, and relational dynamics users seek in a social media space. In contrast to technical descriptions or feature specifications, spatial metaphors draw on familiar experiences~---~such as ``a pajama party at 4am'' or ``a wedding dinner party with invited friends and family''~---~to convey complex affective and social qualities. 

Each metaphor may, explicitly or implicitly, encode several interrelated dimensions of a social space, ranging from the general atmosphere (e.g., lively, peaceful, intimate) to the setting (rooftop, alleyway, festival ground), to the time of day and the type of people who are in attendance. To capture these details, participants in our user study (described in full detail in Section~\ref{sec5:study-design}) provided a metaphor similar to the examples above, which our system elaborated upon and populated an interactive simulated social media platform. The system we created for this study operates in three main phases: (1) translation of the provided metaphor into a discrete set of features, as described in \ref{sec3.1:metaphors-to-features}, (2) the generation of an interactive social platform based on the selected features, as described in Section \ref{sec3.2:populating-system}, and (3) populating the system with social agents derived from the social context of the metaphor, as detailed in Section~\ref{sec3.3:social-agents}. We provide a summary diagram of the system and the study design in Figure~\ref{fig:system-study-overview}.

%\begin{itemize}[nosep]
%    \item The atmosphere or emotional tone of the space (e.g., lively, peaceful, intimate)
%   \item The type of place or setting it takes place in (e.g., rooftop, alleyway, festival ground)
%    \item The time the action is taking place (e.g., early morning, late night)
%    \item The kinds of people or interactions you imagine happening in that space
%    \item How people gather, linger, or move through the space (e.g., spontaneous clustering, sequential rotation, persistent presence).
%\end{itemize}

%To translate metaphorical user input into a functional social media simulation, our system follows a two-step pipeline: (1) extracting core social attributes from spatial metaphors, and (2) mapping those attributes to functional platform features. The result is a partial set of features drawn from our defined functional taxonomy, enabling the modular composition of social environments grounded in user-imagined metaphors.

\subsection{Converting a Metaphor into Social Attributes}\label{sec3.1:metaphors-to-features}

\subsubsection{Translating metaphor into structured description}
The conversion process begins by translating the user’s input~--~the spatial metaphor~--~into a structured template. The template expands the metaphor into eight distinct attributes, shown in Table~\ref{tab:attributes} that specify the underlying social dynamics. The purpose of this step is to clarify the metaphor before translating into a feature set, as early system testing found that direct translation from the metaphor input to a feature list resulted in outcomes that felt random or disconnected from the initial metaphor. An LLM prompt, provided in the supplementary materials, was used to perform this translation.

The structured template reads as follows:

\begin{quote}
``In a space that feels \textit{[atmosphere]}, people come together \textit{[reason for gathering]}, often connecting \textit{[connection style]}. They usually \textit{[duration of participation]}, interact through \textit{[communication style]}, and present themselves using \textit{[identity type]}. Most people are here to \textit{[interaction goal]}, and they have the option to \textit{[control over participation or visibility]}.''
\end{quote}

Each element of the template corresponds to a social attribute as shown in the Table~\ref{tab:metaphor-attributes}:

\begin{table*}[h]
\centering
\caption{Eight metaphor attributes and their design interpretations}
\label{tab:attributes}
\label{tab:metaphor-attributes}
\begin{tabular}{p{3.5cm} p{10cm}}
\toprule
\textbf{Attribute} & \textbf{Description} \\
\midrule
\textbf{Atmosphere} & Emotional and sensory qualities of the space, shaped by spatial density, proximity, and design. Examples include intimate, vibrant, or warm environments. \\

\textbf{Reason for gathering} & The reason people come together—either thematically (e.g., shared interests, events) or relationally (e.g., friends, families, support networks). \\

\textbf{Connection style} & How the space facilitates connection, such as through shared routines or overlapping presence. Integrated spaces afford spontaneous or repeated interaction. \\

\textbf{Duration of participation} & Duration and frequency of participation—ranging from drop-in or episodic interaction to long-term or recurring presence. \\

\textbf{Communication style} & The dominant interaction style, such as reciprocal dialogue, one-to-many broadcasting, or asynchronous messaging. Influenced by spatial layout and social norms. \\

\textbf{Identity type} & The identity mode participants adopt—public, pseudonymous, or anonymous—which affects role visibility and expression. \\

\textbf{Interaction goal} & The primary focus of communication, whether task-oriented, informational, or focused on relationship-building and social cohesion. \\

\textbf{Control over participation or visibility} & The degree of user agency over engagement, from passive observation to selective or continuous participation. Shapes the threshold for interaction. \\
\bottomrule
\Description{The table shows the eight metaphor attributes and their design interpretation descriptions on the right, listing atmosphere, reason for gathering, connection style, duration of participation, communication style, identity type, interaction goal, and control over participation or visibility as the eight attributes.}
\end{tabular}
\end{table*}

These eight attributes serve as intermediate representations between user imagination and the concrete system features of the simulated social platform. We treat user-generated metaphors not merely as descriptions of abstract spatial configurations, but as expressions of \textit{place}~--~spaces imbued with social meaning, practice, and expectation~\cite{harrison1996re}. 
The attributes were derived to reflect dimensions essential to shaping social interaction: affective tone, temporal structure, communicative flow, identity configuration, and the regulation of participation. 
These capture the relational and experiential aspects through which a social space acquires meaning for its participants. 
% Furthermore, they were selected to support the transformation of metaphorical input into multiple levels of platform design features.

% \paragraph{Step 2: Feature Hierarchy Construction}
% Then, the extracted attributes are translated into a concrete list of platform features used to instantiate the interactive simulation environment. These features are organized across a three-level hierarchy to reflect different layers of social system design:

% \begin{itemize}[nosep]
%     \item \textbf{[LV1] Network Structure:} Includes feed structure (e.g., chronological vs. topic-based), content prominence, and comment threading logic.
    
%     \item \textbf{[LV2] Interaction-Level Features:} Elements such as messaging style (e.g., public vs. private), visibility of posts, or types of reactions enabled.
    
%     \item \textbf{[LV3] Advanced Features:} Higher-level constructs such as participant roles, space boundaries (e.g., invite-only vs. open access), and community norms or participation thresholds.
% \end{itemize}

% This hierarchical conversion allows the system to compose a dynamic social simulation that not only reflects the surface metaphor, but also captures the interactional and architectural implications embedded in user imagination.

%\subsection{Step 2: Mapping Social Attributes into Platform Features}

\subsubsection{Translating structured description into platform features}
After extracting the eight metaphor attributes described in the previous section, our system translates these into a structured set of social platform features that define the simulation environment. To support a broad range of imagined social interactions, we enumerated a core set of social media features based on the analysis of real-world social media platforms. 
The feature set enables the modular composition of platform configurations that can be tailored to different types of spatial metaphors, enabling meaningful recombination of design elements. 
This translation pipeline forms the foundation of our metaphor-to-simulation process by systematically mapping conceptual dimensions (e.g., ``a cozy dinner with friends'') into the mechanics of social media interaction (e.g., group-based messaging, real names, ephemeral content).

To build a comprehensive and representative feature set, we analyzed 15 popular social media platforms—such as Instagram, Reddit, WhatsApp, and Discord—that reflect diverse interaction paradigms, community structures, and feature sets. Platform selection was informed by prior comparative work on social systems~\cite{zhang2024form}, and refined through our focus on platforms that prioritize user-to-user interaction. Note that we do not claim that this feature set includes all possible social media features by any means, but rather that it includes a representative set of features currently used on major social platforms. We discuss the limitations of this approach in Section~\ref{sec6.2:future}.

We selected platforms based on the following criteria:
\begin{itemize}
    \item Platforms that emphasize interpersonal and group-based interactions rather than passive content consumption;
    \item Platforms that support reciprocal or multi-party exchanges (e.g., messaging, commenting, group chats);
    \item Platforms that primarily use text as a dominant communication modality.
\end{itemize}

%We excluded platforms such as YouTube and Twitch that center on one-directional, creator-to-audience broadcasting, as their interaction structures do not reflect the metaphorically rich, participatory environments our system aims to support.

\noindent The feature set was developed through a three-stage process:

\begin{itemize}
    \item \textbf{Stage 1: Feature Inventory and Documentation} Two researchers conducted systematic interface walkthroughs across selected platforms to identify and document user-facing UI elements central to social functionality. This included features related to posting, messaging, reacting, and managing social connections. We excluded features related to monetization, analytics, or developer tools in order to maintain focus on user-level interaction components.
    \item \textbf{Stage 2: Functional Analysis} Each feature was then analyzed in terms of its functional role within the broader social system. We categorized features according to their contributions to interaction type (e.g., 1:1 messaging, group discussion), feedback (e.g., likes, upvotes), content flow (e.g., flat vs. nested threads), and identity representation (e.g., anonymous vs. public profiles). This analysis allowed us to understand how platform elements shape interaction dynamics.
    \item \textbf{Stage 3: Hierarchical Organization}
Finally, features were organized into a three-level hierarchical taxonomy based on their functional scope--from structural components to fine-grained interaction controls. This hierarchy enables our system to flexibly combine design elements in a way that reflects both the structure and atmosphere of user-generated metaphors.

\end{itemize}

\noindent The output of this phase of the system is a feature set that supports social interactions that are conceptually aligned with the initial spatial metaphor input. A full description of the set of features used in this system is available in Appendix \ref{app-features}. %This set is drawn from a functional taxonomy organized into three hierarchical levels, each representing a distinct layer of social media design—from foundational structures to fine-grained user controls. Together, these features translate user-imagined metaphors into concrete interaction mechanics within the simulated platform.

%This taxonomy enables us to configure simulated environments that reflect both the high-level structure and fine-grained dynamics implied by users’ spatial metaphors. By grounding metaphorical imagination in a flexible and modular feature system, we enable end-to-end translation from conceptual design to interactive social experience.

\subsection{Generation of an Interactive Social Platform}\label{sec3.2:populating-system}
% This simulation is composed of two core elements: (1) an interactive sandbox interface that assembles UI components based on the selected feature set, and (2) LLM-driven social agents that engage in contextually grounded interactions derived from the user’s metaphor. In this section, we detail the composition of the sandbox environment and describe the architecture behind agent behavior, illustrating how the system brings metaphor-based social dynamics through interactive simulation.
In this section, we detail the composition of the simulation sandbox environment. Our system is implemented using a full-stack architecture. The frontend is developed in React.js, and the backend is built with Node.js, and uses SQLite as the relational database for storing user sessions, configurations, and simulation status. Social agents are powered by OpenAI’s GPT-4o model, which generates real-time responses based on each agent’s predefined role, behavioral traits, and the metaphor-informed context of the simulated environment. 
% To generate agent behaviors, we integrate OpenAI's GPT-4o model, which produces real-time, contextually grounded responses aligned with each agent.
% Our system is implemented using a full-stack architecture. The frontend is built with React.js, enabling modular, component-based rendering of UI elements based on the metaphor-derived feature set.  The backend is built with Node.js, which handles API routing, interaction logic, and data management. SQLite serves as the relational database for storing user sessions, selected feature configurations, and simulation states.

Building on the metaphor-to-feature conversion pipeline described in Section~\ref{sec3.1:metaphors-to-features}, our system instantiates a dynamic and interactive social media environment that reflects the structural and experiential aspects of the user’s imagined space. To populate the social platform with agents that will be later discussed in Section~\ref{sec3.3:social-agents}, we built an interactive sandbox interface that assembles UI components based on the selected feature set. This process leverages the three-level feature taxonomy to guide the hierarchical construction of the interface.

% To instantiate user-imagined social spaces as functional prototypes, our system composes a dynamic UI sandbox based on the structured feature set produced by the metaphor conversion pipeline. This process leverages the three-level feature taxonomy to guide the hierarchical construction of the interface, ensuring that both the structural logic and experiential quality of the user’s metaphor are accurately represented in the resulting simulation.

\paragraph{Level 1: Layout Structure}  
At the highest level, the system determines the macro-structure of the interface--specifically whether the environment should operate through a \textit{feed-based} or \textit{chat-based} layout, and whether the social connections are \textit{group-based} or \textit{network-based}. This is implemented by conditionally iterating through a combination of several predefined layout templates. 
% For instance, a feed-based layout renders a scrollable content stream component, while a chat-based layout uses a flexible message pane with real-time socket connections to simulate continuous dialogue. 

\paragraph{Level 2: Component Composition}  
Once the layout is established, the system populates the interface with core interaction components drawn from the Level 2 taxonomy.  These include messaging windows, post threads, reaction buttons, and identity markers. The configuration of these components--such as flat vs. nested commenting, or real-name vs. pseudonymous display--is modularly designed, allowing different combinations to represent diverse social interaction patterns.

\paragraph{Level 3: Element Configuration and Behavioral Logic}  
At the most granular level, the system adjusts individual UI behaviors and backend logic according to Level 3 features. These include the visibility scope of content (e.g., public vs. private posts), content lifetime (e.g., ephemeral content), and discovery algorithms.  For instance, ephemeral content is managed through timestamp-based expiration and content recommendations are filtered based on the agent’s interest tags and proximity to the user's activity context.
\\
\\
Together, this compositional approach allows the UI to reflect the layered nature of social metaphor: from broad spatial structure to fine-grained social dynamics. Each simulated space thus operates not as a static template, but as a dynamically configured environment shaped by the metaphorical logic.

\subsection{Populating LLM-Driven Social Agents Derived from User Metaphors}\label{sec3.3:social-agents}

Our system incorporates LLM-driven social agents that engage in contextually grounded interactions based on the user’s metaphor. In this section, we describe the architecture underlying agent behavior, including: (1) populating agents and constructing network relations between them, and (2) implementing social activity through three types of behavioral traits--activity, engagement, and updating behavior dynamics. We illustrate how the system uses these components to populate real-time interacting agents and simulate evolving social dynamics.

% To generate agent behaviors, we integrate OpenAI's GPT-4o model, which produces real-time, contextually relevant responses aligned with each agent’s assigned traits and roles.

\subsubsection{Populating Social Agents and Network Construction}
Each agent is initialized with a social role that defines its interaction goals and behavioral tendencies. These roles are adapted from prior research on user behavior in social media environments~\cite{bechmann2013mapping}, capturing a range of participation styles from active contributors to passive observers. Table~\ref{tab:agent-roles} summarizes the agent roles used in our simulation.

\begin{table}[h]
\centering
\caption{Social roles and primary goals for LLM agents}
\label{tab:agent-roles}
\begin{tabular}{ll}
\toprule
\textbf{Role} & \textbf{Goal} \\
\midrule
Influencer & Gain followers and increase visibility \\
Spreader & Disseminate ideas or information \\
Support-Seeker & Find emotional support or affirmation \\
Entertainer & Engage others with humorous or creative content \\
Moderator & Facilitate and regulate discussions \\
Activist & Raise awareness of social or political issues \\
Networker & Connect with like-minded individuals \\
Lurker & Observe without active engagement \\
Bully & Disrupt or provoke with hostile comments \\
\bottomrule
\Description{The table shows the nine social roles and their primary goals for the LLM agents. The roles include influence, spreader, support-seeker, entertainer, moderator, activist, networker, lurker, and bully, with respective goals on the right.}
\end{tabular}
\end{table}

Agents are further customized using a set of behavioral traits that modulate their interaction tendencies. These traits include: (1) \textbf{activity traits}, such as the likelihood of posting, commenting, or reacting; (2) \textbf{engagement traits}, such as messaging or responding to notifications; and (3) \textbf{update behaviors}, such as adjusting relationships or content visibility. Each trait is assigned a probability value (ranging from 0 to 1), which governs how frequently the agent engages in a corresponding activity during simulation.

\begin{table*}[h]
\centering
\caption{Agent Actions Grouped by Behavioral Traits}
\label{tab:agent-actions}
\renewcommand{\arraystretch}{1.2}
\begin{tabular}{
  >{\centering\arraybackslash}p{4.2cm}
  >{\centering\arraybackslash}p{4.2cm}
  >{\centering\arraybackslash}p{4.2cm}
}
\toprule
\textbf{Activity Traits} & \textbf{Engagement Traits} & \textbf{Update Behaviors} \\
\multicolumn{1}{c}{\textit{\small Content creation and direct interaction}} &
\multicolumn{1}{c}{\textit{\small Messaging and communication engagement}} &
\multicolumn{1}{c}{\textit{\small Changes to relationships and visibility}} \\
\midrule
Add post & Start new chat & Send friend request \\
Add channel post & Start new group chat & Accept friend request \\
Add an ephemeral content & Send message in 1:1 chat & Update relation \\
Add comment on post & Send message in group chat & Update restriction \\
Add comment on comment & Create new channel & Update post visibility \\
React & Join channel & \\
 & Read unread messages & \\
\bottomrule
\Description{The table shows agent actions grouped by three behavioral traits of activity traits, engagement traits, and update behaviors. Activity traits refer to content creation and direct interaction, with the respective action items below. Engagement Traits refer to messaging and communication engagement, with the respective action items below. Update behaviors refer to changes to relationships and visibility, with respective action items below.}
\end{tabular}
\end{table*}

In addition, each agent is assigned the following attributes by the system:
\begin{itemize}[nosep]
    \item \textbf{Topical Interests:} At least three topics derived from the user’s metaphor content orientation (e.g., art, politics, health).
    \item \textbf{Personality:} A pseudonymous label, such as \textit{CuriousNeighbor}, constructed from the metaphor’s actor type and connecting environment, guiding the tone and expression of the agent’s interactions.
\end{itemize}

These characteristics are embedded into the system’s LLM prompting schema, ensuring that each agent exhibits consistent behavior across different interaction modes such as posting, replying, reacting, and messaging. By combining social roles, behavioral traits, and metaphor-informed attributes of social space, our system simulates a population of agents that reflects the diversity of real-world social motivations and participation styles.

The overall number of agents generated depends on the metaphorical context. For example, a metaphor that implies a large audience---such as a ``concert''---will result in the creation of a larger number of agents, whereas a metaphor describing a tightly connected space---such as ``a small group of friends having a picnic''---will generate only a small number of agents to match the context. For the purpose of maintaining a reasonable generation time during the user study, the maximum number of agents was capped at 100. After the agents are created, the system generates a social network by establishing relationships among them. These relationships include both bidirectional (i.e., friends) and unidirectional (i.e., followers) connections, and are randomly assigned among the generated agents.

\subsubsection{Social Activity}
Once the agents and network structure are fully established, the agents begin to engage in various social activities in the simulated social media space to emulate realistic social media behavior. These activities include posting content, commenting on posts, initiating direct messages, updating personal settings (e.g., post visibility), and reacting to posts through likes or other engagement forms. Each agent's participation level is governed by its behavioral traits, which influence how frequently it engages and which types of interactions it prioritizes.

To capture the complexity of real-world social media environments, we define and implement 18 distinct agent actions. These actions reflect the breadth of user engagement patterns, ranging from content creation and direct messaging to passive observation and network-building behaviors. Each action is categorized according to one of the three behavioral dimensions in Table~\ref{tab:agent-actions}, corresponding to the agent’s behavioral profiles: activity traits, engagement traits, and update behaviors.

To introduce realistic temporal dynamics, the simulation models each agent’s activity by evaluating both their online presence and their likelihood of participating at a given moment. When generating an agent action, the system begins by checking whether an agent is considered ``active'' in the current simulation cycle. This decision is based on the agent's predefined activity level, which influences their probability of becoming active or remaining idle. Agents with higher activity levels are more likely to enter or stay in an active state, whereas less active agents may be marked as offline or passive.

Once an agent is deemed active, the system determines what action they will take next. This begins with identifying all possible actions available within the current simulation environment—filtered by interface constraints (e.g., agents won’t be able to create new channels in environments without channel-based structures). Then, the system evaluates which actions best fit the agent's individual behavioral tendencies by applying a weighting function based on their behavioral traits. These weights represent how strongly an agent is inclined to perform specific types of actions, such as messaging, posting, or reacting. From this weighted list, final actions are determined probabilistically. As the full implementation of these actions required many pages of prompts, we present in-depth implementation details for each of these behaviors in the supplemental materials.
% The prompts used in the following implementation are listed in Appendix~\ref{}.

% \input{doc/3. Metaphorical Social Design}
% \input{doc/4. Simulation Pipeline}

\section{Study Design}\label{sec5:study-design}
To examine the system's ability to  generate diverse social spaces through metaphor conversion, we conducted a two-phase study involving contrasting types of social spaces: \textit{open social spaces} and \textit{closed social spaces}. These conditions reflect varying degrees of \textit{contextual openness}, defined as the extent to which a social environment feels accessible, loosely structured, or selectively bounded. We selected these two types of social structures because they represent a key design tension in the conceptualization of social media spaces~\cite{gibbs2013overcoming}; open spaces are typically perceived as public, spontaneous, and broadly accessible, while closed spaces tend to be more private, intimate, and selective in participation. Social media platforms can also be divided into more open or closed spaces. For example, most Discord servers would be considered more closed spaces, with invite-only participation and smaller numbers of users in each space, while Twitter (X) would be considered a more open space. Contextually open and closed spaces are roughly aligned with network-based and group-based platforms, though there are exceptions; a publicly-accessible Discord server with hundreds of thousands of members may technically be a group-based space, but functionally it feels much more open. Each participant was asked to imagine an example of each type, with the order randomized across participants to mitigate any order effects.

\subsection{Participants}

We recruited 32 participants from a university community. Participants ranged in age from 18 to 34 years (M = 23.5). Eligibility criteria included regular experience with social media and the ability to imagine a type of social media space they would personally like to use. All participants provided informed consent prior to participation, as approved by the university's Institutional Review Board (IRB). Each participant received 25,000 KRW (approximately \$18.26 USD at the time of the study) in compensation for completing the session, which lasted approximately 1 to 1.5 hours.

\subsection{Procedure}
For each type of social space, participants completed a four-phase process using the system. First phase, participants were prompted to articulate a social media space they would like to use by describing it through a spatial metaphor. This step was designed to capture their imagined social environment in an open-ended and creative manner. After submitting a metaphor, the system was launched in the background to begin constructing the simulation, as the simulation generation process took approximately five to ten minutes depending on the number of simulated users.

In the second phase, while the simulation was populating in the background, participants elaborated on their metaphor using a structured metaphor template that included eight key attributes utilized by our system’s metaphor conversion pipeline: atmosphere, place type, temporality, actor type, interaction flow, content orientation, participation control, and openness. The purpose of this phase was to see whether the system's detailed interpretation of the metaphor was aligned with the participant's. 

In the third phase, participants were presented with a system-generated set of platform features derived from their metaphor input. They reviewed each feature and completed a survey evaluating how accurately the features reflected their original metaphor, using a five-point Likert scale (ranging from ``Strongly disagree'' to ``Strongly agree''). Participants were also asked to identify the top five features that felt most aligned or most mismatched with their metaphor and to provide justifications for their evaluations in open-ended form. This phase allowed us to assess the fidelity of the metaphor-to-feature conversion process.

In the final phase, participants entered a simulated social media environment instantiated based on the feature set generated from their metaphor. As described in a previous section, this environment was populated by LLM-driven agents that responded to user actions in real time, simulating social interactions such as posting, commenting, reacting, messaging, and networking. Participants were encouraged to engage freely with the space for approximately 10–15 minutes per condition. After exploring the environment, they completed a post-survey reflecting on their interaction experience. Specifically, they evaluated the alignment of various interaction modalities~--~including content, comments, reactions, messages, and perceived social connections~--~with their original metaphor. Following the survey, a short semi-structured interview was conducted, during which participants reflected on metaphorical coherence, emotional resonance, and the perceived quality of agent interactions. At the conclusion of the session, participants were asked to compare their experiences across the open and closed space conditions, focusing on the system elements that contributed to distinctions between the two types of social spaces. This comparison aimed to evaluate the system's ability to simulate different types of spaces.

\subsection{Data Collection and Analysis}
A range of quantitative and qualitative data was collected from each session, including participants’ metaphor descriptions and structured templates, their survey responses evaluating system-generated feature alignment and interactions within the simulated environments (e.g., posts, replies, messages, and agent reactions). We also collected open-ended reflections after each simulation and final comparative reflections between the two conditions.

To analyze the data for \textbf{RQ1}~--~How people conceptualize digital social spaces through metaphor~--~we conducted qualitative coding of the metaphor descriptions to identify recurring patterns in how participants conceptualized social spaces and what dimensions they emphasized. We focused on four interpretive dimensions that have been used to explain how people relate to and make sense of places: \textit{Affect}, \textit{Place Orientation}, \textit{Temporality}, and \textit{Social-Interpersonal Importance}~\cite{altman2012place}.

\vspace{0.5\baselineskip}
\begin{itemize}[nosep]
    \item \textbf{Affect (Cognition / Practice):} What emotions and mental or embodied activities structure the imagined space?
    \item \textbf{Place Orientation:} How specific, scaled, or abstract is the imagined environment?
    \item \textbf{Temporality:} What durations shape how people imagine time in social spaces?
    \item \textbf{Social-Interpersonal Importance:} Who is imagined to be present, and what social bonds or dynamics matter?
\end{itemize}
\vspace{0.5\baselineskip}

We conducted inductive thematic analysis on each metaphor along the identified dimensions, referring to the human-authored descriptions to identify which aspects users emphasized through their spatial language.

% To address \textbf{RQ2}, we compared the human-authored metaphor descriptions with the system-generated descriptions for each of the eight template attributes. This comparison allowed us to evaluate the effectiveness of our metaphor conversion pipeline and to identify areas of alignment or conflict. Given both the difficulty of establishing a consistent standard for identifying conflicts across all attributes and the scale of the data—512 pairwise comparisons (8 attributes $\times$ 32 participants $\times$ 2 conditions), we employed a large language model (LLM) to assist in the comparison process. 
% We defined a \textit{conflict} as a case in which the human and system descriptions emphasized substantially different interpretations or social functions for the same attribute, indicating that the LLM pipeline had interpreted the metaphor in a way that was meaningfully different from how the user had imagined it, potentially leading to an inaccurate simulation. Minor differences in tone, phrasing, or elaboration were not considered conflicts. The full prompt and example outputs used for conflict evaluation are provided in Appendix~\ref{appendix:llmconflict}.

For \textbf{RQ2}~--~What features are key to blueprinting a user's imagined social space into a functional platform~--~we analyzed participants’ survey responses regarding the alignment of system-generated features with their original metaphor. We examined which features were consistently rated as highly aligned or misaligned and reviewed participants’ open-ended responses to identify which dimensions of metaphor were most influential in shaping perceptions of feature fidelity.

For \textbf{RQ3}~--~How users' experiences in the simulated environments align with their expectations~--~we used both post-survey responses and interview data to evaluate the experiential quality of the simulated environments. We compared alignment scores across open and closed spaces and analyzed participants’ reflections to understand which design elements contributed to the sense of coherence or dissonance with their metaphor.  
This mixed-methods approach allowed us to assess how well the system  instantiated users’ metaphors and to identify which types of spaces and features were better supported by the system.

\section{Study Results}
In this section, we present findings from our user study examining how participants conceptualized social media spaces through metaphor, how effectively the system translated these metaphors into functional simulations, and how users evaluated the resulting experiences. Our analysis follows four stages: characterizing user-imagined metaphors, assessing system alignment with user intent, identifying key platform features that influenced perceived fit, and analyzing how users judged the authenticity of the simulated social environments. 

\subsection{RQ1: How Do Users Conceptualize Social Spaces Through Metaphor?}
The first research question examines how individuals conceptualize digital social spaces through metaphor, and which metaphorical dimensions are emphasized differently when imagining \textit{open} versus \textit{closed} social environments. In order to usefully simulate digital social spaces, a system must be able to accurately capture substantively different types of spaces, and this research question helps us identify what kinds of different spaces we should aim to be able to simulate. Our study identified a diverse range of metaphorical expressions that participants used to articulate their idealized social settings, shaped by affective tone, spatial openness, and interpersonal dynamics.

\textit{Open space metaphors} often depicted expansive, loosely structured, and socially vibrant environments where a wide range of participants could engage with minimal barriers. Participants commonly referenced public or semi-public gatherings, such as ``a playground'' or ``a job fair'' These metaphors foregrounded themes of spontaneity, ambient connection, and informal social exploration. Some metaphors evoked formalized settings~--~such as ``a debate stage for election candidates''~--~yet were still considered open due to their emphasis on public visibility and performative self-expression. Additionally, bounded social interactions occurring in open spatial settings, e.g., ``chance meetings at the shop/supermarket'' were also coded as open spaces. This demonstrates how participants conceptualized the spatial affordances of openness and accessibility, even when individual interactions are brief or transient.

In contrast, \textit{closed space metaphors} described more intimate and selectively accessible environments, often characterized by emotional closeness, mutual trust, and a sense of privacy. Examples include ``a neighborhood bar,'' ``my bedroom with close friends,'' and ``a homeparty with cozy atmosphere and delicious foods.'' %\footnote{Note that the term ``homeparty'' (집들이) used here has a Korean cultural connotation that is significantly different from the commonly-used Western term ``house party,'' indicating a much smaller, more intimate and personal event, somewhat similar to a small housewarming party.} 
These metaphors reflect participants’ desires for emotionally safe, bounded settings that support sustained engagement and deeper interpersonal relationships. However, some metaphors--such as ``a mosh pit at a Travis Scott concert''--were categorized as closed spaces despite their large public settings. In this case, the metaphor conveyed a tightly focused, immersive experience centered on emotional intensity and collective bonding, suggesting a closed social atmosphere within a broader open space.
%The full list of participant-generated metaphors is presented in Appendix~\ref{}.

As evident from the diverse metaphors described above, participants drew on a variety of perceptual cues and experiential references to articulate what made a space feel open or closed. To systematically investigate the underlying patterns in these interpretations, we employed a four-dimensional coding scheme derived from spatial and environmental psychology, which identifies key interpretive dimensions through which individuals relate to and make sense of place~\cite{altman2012place}. These four dimensions~--~\textit{Affect}, \textit{Place Orientation}, \textit{Temporality}, and \textit{Social-Interpersonal Importance}~--~were used to code each metaphorical description. 

\textit{Affect} captures the emotional tone of the space (e.g., ``Energetic,'' ``Relaxed,'' ``Tense''); \textit{Place Orientation} reflects the social function or spatial focus (e.g., ``Activity-centered,'' ``Spectator-based,'' ``Identity-expressive''); \textit{Temporality} refers to time-based structuring of interaction (e.g., ``Spontaneous,'' ``Long-Stay,'' ``Episodic''); and \textit{Social-Interpersonal Importance} accounts for how relationships and participation are structured (e.g., ``Peer-driven,'' ``Opt-in,'' ``Goal-oriented'').
Table~\ref{tab:dimension-distribution} presents the frequency distribution of coded labels across open and closed metaphors. %Full label definitions are provided in Appendix~\ref{appendix:dimension_labels}.

\begin{table}[h]
\centering
\caption{Distribution of interpretive dimension labels across open and closed space metaphors}
\label{tab:dimension-distribution}
\begin{tabular}{lcc}
\toprule
\textbf{Label} & \textbf{Open} & \textbf{Closed} \\
\midrule
\multicolumn{3}{l}{\textit{Affect}} \\
\quad Energetic & 9 & 3 \\
\quad Relaxed & 5 & 14 \\
\quad Playful & 5 & 7 \\
\quad Welcoming & 9 & 2 \\
\quad Engaged & 14 & 7 \\
\quad Intimate & 1 & 11 \\
\quad Tense & 6 & 4 \\
\multicolumn{3}{l}{\textit{Place Orientation}} \\
\quad Activity-centered & 6 & 6 \\
\quad Spectator-based & 7 & 2 \\
\quad Identity-expressive & 4 & 4 \\
\quad Social bonding & 13 & 17 \\
\quad Knowledge-exchange & 9 & 5 \\
\multicolumn{3}{l}{\textit{Temporality}} \\
\quad Spontaneous & 13 & 3 \\
\quad Scheduled & 5 & 4 \\
\quad Long-Stay & 8 & 17 \\
\quad Episodic & 5 & 7 \\
\multicolumn{3}{l}{\textit{Social-Interpersonal Importance}} \\
\quad Peer-driven & 5 & 8 \\
\quad Low visibility & 4 & 4 \\
\quad Exchange-driven & 8 & 6 \\
\quad Opt-in & 7 & 8 \\
\quad Goal-oriented & 7 & 5 \\
\bottomrule
\Description{The table shows the distribution of the four interpretive dimension labels across open and closed space metaphors, with the open on the left and closed on the right. The four interpretive dimensions are affect, place orientation, temporality, and social-interpersonal importance, with respective items below.}
\end{tabular}
\end{table}

The analysis reveals clear contrasts along the dimensions of \textbf{Affect} and \textbf{Temporality}. Open space metaphors were frequently associated with affective tones such as \textit{Energetic}, \textit{Welcoming}, and \textit{Engaged}, reflecting dynamic and socially vibrant environments that encourage interaction among diverse participants. In contrast, closed space metaphors emphasized more \textit{Relaxed} and \textit{Intimate} moods, suggesting emotionally secure and personally meaningful settings designed for deeper, more sustained engagement.

For the \textbf{Temporality} dimension, open metaphors leaned toward \textit{Spontaneous} interactions, highlighting momentary or serendipitous encounters. Closed metaphors, however, predominantly reflected \textit{Long-Stay} temporalities, indicating spaces where participants envisioned prolonged, continuous interaction with familiar others. Together, this dimension-based analysis suggests that metaphors for digital social spaces—particularly those distinguished by contextual openness—highlight how users rely on contrasting interpretive dimensions to express different modes of social interaction. A successful system for social media simulation must be able to output different spaces that clearly fall within distinct parts of this dimensional space, ranging from spontaneous, energetic spaces to slower, more intimate spaces.

\subsection{RQ2: What System Features Effectively Represent User-Imagined Spaces?}
Where RQ1 focused on users' mental concepts for social spaces, RQ2 explores the design of \textit{digital} social spaces on a much more granular level, identifying which key social features are necessary to blueprint a user's imagined social space into a functional social platform. Drawing on survey data, we identified the top five features most frequently perceived as aligned with participants’ envisioned environments for both open and closed space types. Our analysis considered both alignment and misalignment patterns to assess which design elements are consistently critical across contexts, as well as which features are distinctly important for supporting the unique interactional dynamics of open versus closed social spaces. Note that, in this context, high alignment for a feature does not mean that the mere \textit{presence} of that feature was aligned with participants' expections; instead high alignment means that the pipeline was consistently able to select and instantiate a \textit{variation} of that feature that met participants' expectations. For example, high alignment in \textit{Identity}, as described below, meant that the system was consistently able to determine whether a space should be real-name, pseudonymous, or anonymous based on a participant's metaphor. 

\begin{figure*}
    \centering
    \includegraphics[width=\textwidth]{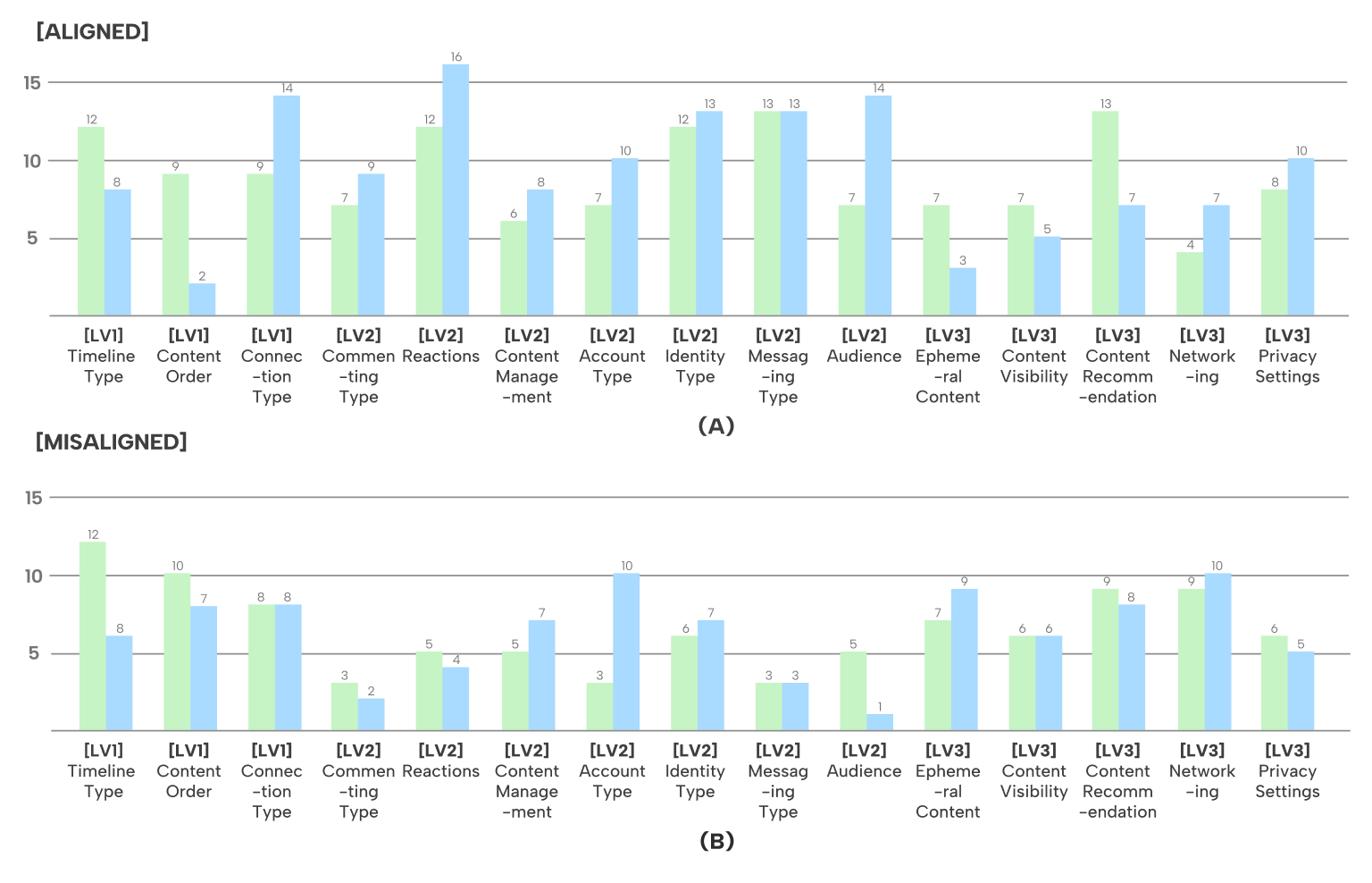}
    \caption{Results of participants' feature alignment ratings for open (green) vs closed (blue) bars. (A) shows the number of participants who identified each feature as aligned with their metaphor. (B) shows the number who identified each feature as misaligned.}
    \Description{The figure shows two graphs of the participant feature alignment rating results. It displays a list of bar graphs for each feature, totaling 15 features. On the left are the open space ratings in green bars, and on the right are the closed space ratings in blue bars. On top is the graph for number of participants who identfied each feature as aligned with their metaphor, and below is the graph for number of participants who identified each feature as misaligned with their metaphor.}
    \label{fig:rq3-results}
\end{figure*}

\subsubsection{Aligned Features}
For both open and closed spaces, three features~--~\textit{Messaging Types}, \textit{Reactions}, and \textit{Identity}~--~emerged as the most consistently aligned elements with participants’ imagined social spaces. \textbf{Messaging Types} were frequently identified as aligned across both open and closed settings, with participants emphasizing the need for both private and group communication. Regardless of a space's openness, users envisioned interaction scenarios in which both one-on-one and group messaging were essential. \textbf{Reactions} were also widely considered aligned across cases, as participants valued diverse and expressive ways to convey emotion, regardless of the space's level of openness or intimacy. \textbf{Identity} was another highly aligned feature, though unlike the others, alignment depended more on context-sensitive calibration. Participants appreciated when identity options (e.g., real-name, pseudonymity, or anonymity) matched the social configuration and relational expectations embedded in the space, aligning identity visibility with interpersonal closeness or professional norms.

Building on these shared alignments, certain features demonstrated stronger salience in either open or closed spaces, reflecting the distinct social expectations embedded in each type of metaphor. In open spaces, \textbf{Timeline Types} and \textbf{Recommendation} emerged as particularly well-aligned due to their support for flexible and thematically driven interaction. This reflected an imagined environment in which individuals navigate open spaces by engaging in spontaneous or topic-centered discussions, often shaped by shared interests. By contrast, in closed spaces, \textbf{Connection Type} and \textbf{Audience} were more frequently aligned. Participants associated these features with affordances for intimacy, privacy, and selective interaction, noting that group-based ties and restricted audience visibility better captured the sensitive, emotionally secure dynamics often envisioned in closed metaphors. In each of the above cases, the pipeline was able to identify the expected social nuances and instantiate variations of the features that were aligned with participants' expectations.

\subsubsection{Misaligned Features}
Among the most frequently misaligned features in metaphor translation were \textit{Recommendation}, \textit{Networking}, and \textit{Connection Type}. \textbf{Recommendation} was commonly misaligned due to the system’s preference for instantiating topic-based content suggestions, whereas participants often expected a combination of topic relevance and popularity. In many cases, participants envisioned social contexts~--~such as a guesthouse party, a picnic, or a casual indoor gathering~--~where trending topics would naturally shape conversation. For the \textbf{Networking} feature, the system frequently assumed full block/mute capabilities, while participants described both open and closed settings where such controls felt contextually implausible or socially inappropriate. Whether in open environments like ``a crowded soccer stadium with fans cheering for the World Cup finals match'' or intimate spaces like ``a space where I can have honest conversations with 30 close friends at my place, completely drunk, at 2 a.m.'', users rejected simplistic instantiations of blocking and muting features, as such actions felt incompatible with both public encounters and trusted relational contexts. \textbf{Connection Type} was also frequently misaligned, as the system tended to infer network-based connections for open spaces and group-based connections for closed ones, while participants expressed more nuanced expectations. Some rejected the idea of networked ties even in open spaces (e.g., ``small talk on the street with a stranger''), whereas others found network-based connections appropriate in loosely structured, open environments (e.g., ``a grass field for picnic in front of a big building''). These misalignments suggest that the system’s interpretation of connection type was insufficiently grounded in the specific relational dynamics conveyed by participants’ metaphors.

Beyond these commonly misaligned features, open spaces also revealed frequent mismatches in \textbf{Timeline Types} and \textbf{Content Order}, arising from the wide range of interaction styles participants envisioned. Some imagined open settings as dynamic, chat-like spaces for spontaneous group exchange (e.g., playgrounds or roundtables), while others pictured them as structured environments with feed-based updates supporting trend discovery (e.g., job fairs or conventions). Similarly, content order expectations diverged between those preferring algorithmic relevance and those valuing chronological transparency. These differences highlight how open metaphors span a broad spectrum of interactional distance and discovery preferences, making them harder to capture with uniform system logic. In contrast, misalignments in closed spaces—particularly around \textbf{Account Type} and \textbf{Ephemerality}—reflected more consistent concerns about privacy and trust. Participants imagined closed environments as emotionally secure and enduring spaces, where features like public accounts or disappearing content conflicted with their expectations for selective, lasting connection.

\subsection{RQ3: How Do Simulated Experiences Match User Expectations?}
% \subsection{RQ4: Evaluating Experiential Quality in Simulated Environments}

Where RQs 1, and 2 focused on hypothetical social environments (RQ1) or proposed sets of features (RQ2), RQ3 investigated participants' experiences actually interacting with a simulated social media platform generated based on their metaphor. For RQ3, we evaluated the experiential quality of user interactions within the simulated environments, focusing on five key interaction components: \textbf{Contents} (Posts and Channels, if applicable), \textbf{Comments/Replies}, \textbf{Reactions}, \textbf{Messages}, and \textbf{Social Connections}.

After interacting with the simulated social media platform for approximately 10-15 minutes, participants rated the alignment of each experience with their original metaphor on a 5-point Likert scale (1 = \textit{Not at all aligned}, 5 = \textit{Extremely aligned}). To analyze the results, we first conducted the \textit{Shapiro–Wilk test} to assess the normality of the distribution for each component. If the data met the assumption of normality, we used a \textit{paired t-test}; otherwise, we used the \textit{Wilcoxon signed-rank test}. Quantitative analysis of user ratings revealed no statistically significant differences in perceived alignment between open and closed cases across five interaction components -- Content (Channels), Comments, Reactions, Messages, and Social Connections (see Table~\ref{tab:rq4_scores}). 

\begin{table*}[h]
\centering
\caption{Mean alignment ratings, standard deviations, and statistical test results for interaction components in open and closed simulations}
\resizebox{\textwidth}{!}{%
\begin{tabular}{lccccc}
\toprule
\textbf{} & \textbf{Content (Channels)} & \textbf{Comment} & \textbf{Reaction} & \textbf{Message} & \textbf{Social Connection} \\
\midrule
\textbf{Open (M (SD))} & 3.31 (1.15) & 3.38 (1.07) & 3.84 (1.17) & 3.56 (1.37) & 3.44 (1.13) \\
\textbf{Closed (M (SD))} & 3.56 (1.29) & 3.63 (1.07) & 3.88 (0.94) & 3.69 (1.28) & 3.56 (1.11) \\
\midrule
\textbf{Shapiro-Wilk (p)} & .115 & .054 & \textbf{.049} & .144 & .138 \\
\textbf{Test Used} & Paired t-test & Paired t-test & Wilcoxon & Paired t-test & Paired t-test \\
\textbf{Statistical Result} & $t(31)=0.86,\ p=.397$ & $t(31)=0.80,\ p=.432$ & $Z\approx -0.10,\ p=.919$ & $t(31)=0.43,\ p=.670$ & $t(31)=0.42,\ p=.680$ \\
\bottomrule
\end{tabular}%
\Description{The table shows the mean alignment ratings, standard deviation, and statistical test results for the interaction components in open simulations. The columns are content or channels, comment, reaction, message, and social connection, respectively. The first row is the corresponding figures for open space, with the mean and the standard deviation. Below is the same for the closed space. Below is the Shapiro-Wilk value and the test used. The final row is the statistical result from the test.}
\label{tab:rq4_scores}
}
\end{table*}

This indicates that the quality of simulation was not quantitatively different across the open and closed conditions at a level we could detect with the sample size used in this study. 

Beyond comparison of the two conditions, in-depth interviews revealed qualitative differences in how participants experienced alignment compared to their expectations depending on the openness or closedness of the simulated environment.  Participants most frequently focused on three of the five interaction components in describing their experiences~--~Content, Messaging, and Social Connection~--~and we discuss their feedback for each of these three interaction components below.

\subsubsection{Contents (Posts and Channels)}
Participants highlighted notable differences in how content circulation shaped the perceived openness or closedness of the simulated environments. In open spaces, content was experienced as more ambient and distributed, with multiple posts appearing simultaneously and covering a broad range of topics. This produced a dynamic, parallel form of interaction that contributed to the perception of a socially active and loosely connected space. As one participant explained, \textit{``In the open space, many people were talking simultaneously… In the closed space, it felt more intimate and small-scale.''} (P30) In contrast, closed spaces featured a slower, more sequential flow of interaction. Posts were often perceived as isolated expressions, reinforcing a sense of individual rather than collective engagement. One participant described this dynamic as resembling a niche or insular community: \textit{``Each post here felt like it came from just one person... it made the space feel more closed, like a niche fan café.''} (P22)

The structure of participation and spatial navigation significantly influenced how users experienced openness in the simulated environments. Participants described feed-based content in open spaces as enabling broad visibility and low-friction engagement, supporting ambient awareness and passive participation at scale. In contrast, chat-based formats in closed environments were perceived as sequential and limiting, restricting the number of participants who could meaningfully engage at a given time. As one participant noted, \textit{``Chat is synchronous... the number of people who can participate is limited. Ten thousand people can’t chat together, but ten thousand people might see my post.''} (P20) Beyond the format of interaction, the presence of the \textit{Community Tab} in open environments reinforced a sense of spatial openness and agency. By allowing users to freely explore segmented spaces, it offered a navigational structure that mirrored metaphorical framings of streets, plazas, or clusters of informal gatherings. As one participant explained, \textit{``In the Community Tab, I could just explore and select things freely... that gave me a sense of agency. This kind of freedom matched what I expected from the open metaphor.''} (P10).

\subsubsection{Message}
Participants’ reflections on the messaging component revealed distinctions in how private and group interactions signaled openness or closedness, not only through content but also through tone, purpose, and relational framing. In open environments, group chats often lacked a defined purpose, which led participants to interpret the interaction as casual and non-committal. Even personal topics felt generalized and broadly addressable, reinforcing a sense of openness. As one participant commented, \textit{``There should be some goal or purpose or aim of each chat. No purpose at all for the group chatting now.''} (P16) Another participant contrasted the two cases by noting that the open version felt more like a general public inquiry, while the closed version carried more emotionally direct, situational dialogue: \textit{``This one [closed] felt more like someone checking in on me personally.. like asking about my situation. The other one was more like general questions anyone could answer, like what shows are good.''} (P1)

Beyond purpose and tone, participants also distinguished open and closed cases based on the coherence of message content. In closed environments, message threads were described as thematically interconnected, often anchored by a shared context such as a hobby or group identity. This continuity contributed to a sense of belonging and mutual recognition. In contrast, messages in open environments were perceived as disjointed and episodic—fitting for a public or loosely structured space. As one participant noted, \textit{``In the closed space, the messages were related—it was like a book club. But here it’s very random... which makes sense for public.''} (P17) These distinctions suggest that coherence and framing play a critical role in how users perceive the social boundaries of conversational spaces.

\subsubsection{Social Connection}

Participants’ perceptions of social connections were closely tied to how visible and persistent others’ identities felt within the environment. In closed spaces, features like viewing profiles or past posts contributed to a sense of familiarity and continuity. Participants felt that being able to see more detailed personal information made the environment feel more like a known group or community, where informal relationships could develop over time. As one participant put it, \textit{``In a private SNS, I would think you should be able to see detailed information like what posts someone has written so far.. But somehow, in my metaphor, this [limited view] felt more appropriate.''} (P1) Others noted that profile visibility wasn’t necessary in these settings because the assumption was already one of familiarity: \textit{``Because the system assumes you might not know this person, it shows you this kind of profile card. But in a closed space, you wouldn’t really need that.''} (P27)

In open spaces, by contrast, participants expected a more fleeting and anonymous form of interaction. Visibility features like profile access or follow/follower systems were often viewed as unnecessary, or even misaligned with the context. These environments were described as feeling more like public spaces for casual encounters--where social ties weren’t meant to persist. One participant explained, \textit{``I don’t think I should be able to see their profile or follow them if it’s open space… I’m never going to meet them again.''} (P32) The same participant described the opposite expectation for closed spaces: \textit{``If it’s more like, you know, a closed space you’re like ready to get to know them… you actually think of the possibility of maybe like hanging out with them after the event.''} Others expressed a preference for minimal connection features in closed spaces to maintain a sense of containment and trust. As one participant simply noted, \textit{``In closed spaces, I don’t think follower/following visibility is necessary.''} (P16) These responses show how expectations around social persistence and identifiability help shape whether a space feels open or closed.

\section{Discussion}
This study investigated how spatial metaphors can serve as a generative lens for imagining and simulating new forms of digital social interaction. The following sections reflect on the strengths and limitations of this approach, suggest system-level improvements, and outline directions for future research.

\subsection{Reflections on Metaphor-Driven Social Design}

Our study reveals that metaphor-driven social design offers a compelling framework for re-imagining online social spaces, while also surfacing key challenges and opportunities for refinement. Below, we reflect on the strengths of this approach, its current limitations, and areas where future iterations could be improved.

\subsubsection{Metaphor-driven design foregrounds users' value-driven expectations} A major strength of metaphor-driven design is its ability to elicit value-driven expectations early in the design process. Rather than iterating on fixed paradigms, users begin with a metaphor that captures their personal ideals of social interaction--whether it be openness, intimacy, spontaneity, or routine. This stands in contrast to how new features are typically introduced on existing platforms today: incrementally and often layered atop legacy designs that constrain rethinking of foundational structures. Particularly in the context of social media, where interactions are deeply relational and culturally embedded, it is difficult to imagine fundamentally new paradigms when design is anchored to current defaults. Metaphors offer users a new way to turn the tables on prevailing design assumptions.

In real-world social life, we move through a diverse array of social spaces, including quiet conversations, celebratory gatherings, collaborative workspaces, each shaped by different rules of interaction, visibility, and emotional tone. Our participants’ metaphors reflected this diversity, and this highlights that no single model of social space fits all interaction needs, yet most social media platforms still operate within a limited and homogeneous design paradigm. Social spaces should be designed to support meaningful, situated, and comfortable interaction, adapted to the different ways users seek social connection, support, expression, or presence. As scholars have argued, online social spaces should be grounded in user-centric values, not just efficiency or engagement metrics.

Our method helps address this design rigidity by giving users a new way to explore and evaluate imagined social spaces. Through metaphor, users can describe complex emotional and structural qualities of interaction that are otherwise hard to articulate using conventional design language. This includes abstract but important values such as ``emotional bonding,'' ``flexibility,'' or ``extended engagement.” By rapidly prototyping these metaphorical spaces and simulating interaction within them, our system offers users an immediate and low-cost means of testing their imagination. Importantly, our system also offers a rare opportunity to try out new social interaction models at a scale and pace not possible on existing social platforms, where user feedback loops are slow and implementation is costly. This opens up broader participation in the design of social systems and expands the possibility space for what digital social platforms could become.

\subsubsection{Metaphors may lack concrete specificity} At the same time, our study surfaced several key weaknesses of metaphor-driven design. One of them is the ambiguity and variability of metaphor interpretation. Even when participants used similar metaphors (e.g., a café, a party), they often held different expectations about what those spaces should afford--such as who could speak, how personal the conversation should be, or how visible participants should be to one another. This led to divergent simulation outcomes that sometimes misaligned with user intent. Our current system relies on a one-time metaphor input to generate simulation features, but this does not support the fluidity or contextual flexibility that real-world metaphors often imply.

In addition, our findings in RQ2 show that the system’s interpretation of metaphor frequently diverged from human intention, especially around less visible attributes like temporal engagement and participation control. While the system handled surface-level features such as atmosphere or interaction style relatively well, it often failed to capture more nuanced social attributes by flattening and generalizing the diversity social dynamics. This suggests that metaphor-to-feature mapping needs to be both more granular and more adaptive, capable of handling multiple interpretations or dynamically refining its output based on iterative user input.

\subsubsection{The effectiveness of a simulated space depends on more than just features} Our evaluation also points to specific areas where system functionality can be improved. In RQ3, we observed that content generation showed the lowest overall average alignment among all interaction components (see Table \ref{tab:rq4_scores}). This is unsurprising, as generating realistic and socially appropriate content, particularly for simulating the subtle emotional tones of real users, is a complex challenge, and humans are naturally highly-attuned to divergences from expectations in communication; in essence, despite recent rapid advancements in LLM quality, a significant number of posts in the simulation still fell victim to a sort of uncanny valley effect for text communication. In closed cases, participants noted that agent-generated content failed to evoke the depth of emotional sharing they had imagined, often lacking intimacy or personal relevance. In open spaces, the content felt too formal and disconnected from everyday conversational norms. Channel names and topic categorizations also felt artificial, and chat interactions were frequently perceived as discontinuous or lacking narrative flow. This is an area where providing training data or more domain-specific generation instructions could lead to future improvements.

Future iterations of this system could also refine the system’s functional taxonomy and expand support for a broader range of content modalities. At present, the platform primarily simulates interaction through text-based chat, which constrains the expressive capacity of certain metaphorical spaces. Several participants noted that this modality limited the realism and depth of their experience, particularly when their metaphors implied visual, auditory, or multimodal interaction. For instance, one participant described ``a basement art studio at night where focused, nerdy individuals gather to quietly work and discuss their drawings.'' In this scenario, the absence of image-sharing or sketch-exchange capabilities diminished the plausibility of the simulation. Addressing this gap will also require the extension of the platform’s feature taxonomy to meaningfully map these modalities.  For example, visual metaphors might necessitate functions such as saving or archiving images, curating personal content collections, or circulating visual artifacts through private or group messaging. Mapping these modality-specific features to users’ metaphor-informed expectations would enhance the system’s ability to support a richer and more diverse range of social interactions.

Together, these insights affirm the potential of metaphor as a design lens, while also highlighting the importance of building systems that can support the complicated, evolving, and richly contextual nature of social interaction. Rather than treating metaphors as static inputs, future iterations of this method should explore interactive and co-adaptive metaphor workflows, enabling designers and users alike to treat metaphors not just as a design prompt but as a framework for participatory social system development.

\subsection{Future Directions}\label{sec6.2:future}
Metaphor-driven social design presents a flexible framework for rethinking how social platforms can be imagined, prototyped, and adapted. While this study demonstrates the feasibility of translating metaphorical concepts into system features, further development is needed to address the complexity and diversity of imagined social interactions. In this section, we outline key directions for expanding this approach, focusing on system scalability, design adaptability, and improved simulation accuracy.

\subsubsection{Simulating Richer Social Complexity}
One promising direction lies in improving how the system supports diverse and complex configurations of social features across varied metaphorical spaces.  While our study focused on a finite set of predefined features (e.g., timeline types, connection types, account types), real-world social configurations often involve intersecting qualities, such as bounded timeframes combined with temporary group formations, or mixed interaction layers operating in parallel. For instance, a participant might imagine a ``convention hall where visitors can attend time-limited sessions while also engaging in casual hallway conversations with rotating groups,'' which blends structured scheduling, transient social bonds, and multiple coexisting forms of engagement.

Given the combinatorial scale of possible feature arrangements, it is impractical to define every valid configuration or preprogram all corresponding agent behaviors for each possibility. Instead, future systems should focus on flexible architectures that support modular composition of social features without relying on fixed templates. Rather than assigning fixed behaviors to predefined feature sets—as the current system does by matching each metaphor to a single static configuration, future implementations could structure agent behavior around higher-level interactional goals or constraints derived from users’ metaphor-informed intentions. For example, instead of specifying that an agent must act differently in ``group'' vs. ``network'' connection types, the system might prioritize behaviors that support spontaneous entry and exit from conversations or manage attention among multiple parallel conversations by adjusting responsiveness based on interactional density. This approach would enable more expressive and adaptable simulations aligned with the nuanced complexities found in users' metaphorical concepts.

\subsubsection{Supporting Dynamic and Adaptive Social Spaces}
Another important direction is enabling social media platforms to support the dynamic modification of social space features during use. While our current system generates a static configuration based on a one-time metaphor input, participants often imagined spaces whose social settings evolve over time—for example, a casual chat that becomes more private, or a public gathering that splits into subgroups. One participant described ``a rooftop workspace that gradually turns into a party,'' while another imagined a convention hall where ``a private ephemeral chat thread emerges between visitors who linger at the same booth.'' These temporal and situational transitions are difficult to predefine, yet they are central to how real social experiences unfold.

Future systems should allow users to revise or extend their metaphor-based designs while the simulation is ongoing. This includes introducing new features in response to emerging interaction patterns, such as enabling an ephemeral private chat after a group discussion begins to feel more personal. Technically, this requires platform infrastructures that support live reconfiguration of social features without disrupting interactional coherence. Interfaces must visually and functionally reflect these evolving conditions, while agents must be capable of adjusting their behaviors in response to updated social parameters. The ability to flexibly modify a space during use~--~while maintaining consistency with user intent~--~would shift metaphor-based social design from a static, generative task to an ongoing, participatory process of exploration and refinement.

\subsubsection{Enhancing Visual and Contextual Representation}
Currently, the system focuses on structural translation from metaphor to feature settings, but future iterations could incorporate richer visual and contextual translations of the metaphor. Adding visual feedback, such as interface layouts, spatial arrangements, or ambient cues that align with the intended social setting, could help users more effectively understand and evaluate the simulated space. For instance, environments might visually reflect the spatial or emotional tone of the metaphor: dark, enclosed spaces for private reflection, or brightly animated, open fields for playful discussion. Prior research in interface and spatial design has shown that visual framing significantly shapes user behavior, emotional engagement, and perceptions of affordance in digital environments~\cite{dourish2001where, gaver2003ambiguity, sengers2006staying}.

The way the interface is organized, including how users are represented on screen, the placement of chat panels, or the openness of navigation elements, can influence how individuals interpret interpersonal proximity, visibility, and conversational flow. Likewise, aesthetic elements such as color schemes and typography contribute to setting expectations and shaping the emotional atmosphere of the space. These visual and spatial cues make abstract concepts like openness, intimacy, or formality more accessible through experience. Incorporating metaphor-sensitive spatial and visual design therefore offers a promising direction for creating simulations that are more interpretable, emotionally resonant, and aligned with users' envisioned social dynamics.

\subsection{Limitations}
While our study demonstrates the potential of metaphor-driven social design for prototyping and evaluating imagined interaction spaces, several limitations must be acknowledged. These span from the variability in user-provided inputs to constraints in system implementation and evaluation design. Below, we outline key areas where the current approach falls short and suggest considerations for future refinement.

\subsubsection{Variability in Metaphor Quality and Framing}
One key limitation stems from the variability in the quality and specificity of participant-provided metaphors. While some users offered rich, grounded metaphors, such as ``a safe playground in the afternoon, with parents or patrols stopping by some times'', others provided more abstract or underdeveloped descriptions, such as ``my bedroom''. Though we allowed participants significant freedom to define the scope and level of detail of the provided metaphor in an attempt to avoid over-constraining their input, this inconsistency affected the system’s ability to translate metaphors into meaningful design components, often resulting in misaligned or vague simulations. In addition, the analytical framework primarily relied on a binary classification of \textit{open} versus \textit{closed} spaces. While analytically useful, this dichotomy failed to capture the complexity of metaphors that incorporated hybrid, transitional, or context-dependent social dynamics.

\subsubsection{Agent Behavior Generation and Evaluation Gaps}
Another limitation involves the underdeveloped evaluation of agent behaviors intended to simulate user interaction. Currently, agent responses are generated through prompt-based scripting, and the system lacks a rigorous framework for evaluating whether this content realistically mirrors real-world social media behavior~--~particularly in terms of conversational coherence, content depth, stylistic nuance, and contextual appropriateness. There is also limited assessment of what constitutes socially plausible interaction, and how to measure the perceived quality or believability of agent dialogue. As a result, the current implementation may fall short of capturing the richness and variability of authentic online social exchanges, especially in complex or emotional scenarios.

\subsubsection{Evaluation Scope and Study Population}
The current evaluation focused primarily on participants’ perceived alignment between their metaphors and the system-generated environments. While this approach provided valuable insight into expectation matching, it did not capture actual longitudinal behavioral patterns within the simulation (e.g., message frequency, depth of interaction, turn-taking dynamics). Consequently, the system’s effectiveness remains untested in terms of whether the metaphorical fidelity would hold up under longer-term engagement and scrunity. Furthermore, the participant sample was relatively small and demographically narrow, which limits the generalizability of the findings, particularly for more diverse user populations such as professional platform designers whose interaction needs and metaphorical frames may differ significantly.

\section{Conclusion}

This paper introduces a novel system that leverages metaphor as a generative lens for designing, prototyping, and simulating alternative social media spaces. By enabling users to express their envisioned social environments through spatial metaphors, the system translates these abstract concepts into structured platform features and populates the resulting environments with LLM-driven agents capable of simulating interaction. Through a four-stage study, we explored how users conceptualize digital social spaces, how well the system captures those concepts, and how users evaluate the resulting simulations.
Our findings reveal that users consistently distinguished between open and closed social space, characterizing open spaces as spontaneous, energetic, and engaged, while describing closed spaces as long-stay, intimate, and relaxed. Although the system often produced coherent representations of these metaphors, recurring mismatches, especially around less visible social dynamics, highlight the challenges of interpreting and operationalizing metaphorical input. Ultimately, users evaluated the authenticity of the simulated environments based on how well the experience aligned with their original expectations for social interaction. These results underscore the potential of metaphor-driven simulation while also highlighting the need for more adaptive systems capable of supporting the richness and complexity of users’ imagined social worlds.

\bibliographystyle{ACM-Reference-Format}
\bibliography{main}

%\newpage
%\input{doc/10. Supplemental}

\newpage
\appendix
\section{Features Taxonomy}
\label{app-features}

The sections below summarize the full set of social media features present in the simulated spaces. 

\subsection{Level 1 (LV1): Network Structure}
This level defines the foundational architectural patterns of social platforms, including content access, navigational flow, and relationship structure. These choices shape how content is presented and consumed by influencing the structure and flow of user interactions. See Table~\ref{tab:feature-taxonomy} for more detail.

\subsection{Level 2 (LV2): Core Interaction Mechanisms}

This level includes primary interaction components that define how users engage, express identity, and form social connections. These mechanisms translate metaphorical attributes like ``communication flow'' or ``identity type'' into specific interaction designs. See Table~\ref{tab:feature-taxonomy-starred} for more detail.

\subsection{Level 3 (LV3): Advanced Features and Customization}

This level includes control and personalization features that shape how content flows, how users manage boundaries, and how social visibility is regulated. These features reflect deeper layers of metaphorical nuance, such as privacy, temporal and selective engagement. See Table~\ref{tab:system-level-features-starred} for more detail.

\begin{table*}[h]
\centering
\caption{Network structure (LV1) feature taxonomy used for configuring simulated social media spaces.}
\label{tab:feature-taxonomy}
\renewcommand{\arraystretch}{1.4}
\begin{tabular}{>{\raggedright\arraybackslash}p{4.2cm} >{\raggedright\arraybackslash}p{4.5cm} >{\raggedright\arraybackslash}p{6.3cm}}
\toprule
\textbf{Feature Category} & \textbf{Type} & \textbf{Description} \\
\midrule

\multirow{2}{=}{\textbf{Timeline Format}\\(Specifies the dominant content display mechanism)} 
  & Feed-based & Posts appear in a scrollable stream; common in public or broad-interest spaces. \\
  & Chat-based & Communication occurs in real-time or topic-specific message threads. \\
\midrule

\multirow{2}{=}{\textbf{Content Order}\\(Determines how content is sequenced)} 
  & Chronological & Users see the most recent posts first. \\
  & Algorithmic & Content is sorted by calculated relevance or popularity. \\
\midrule

\multirow{2}{=}{\textbf{Connection Type}\\(Defines the basic relational structure)} 
  & Network-based & Users follow or connect with others to receive content (e.g., Instagram). \\
  & Group-based & Content is visible within shared spaces or communities (e.g., Discord). \\
\bottomrule
\end{tabular}
\end{table*}

\begin{table*}[h]
\centering
\caption{Interaction mechanism (LV2) feature taxonomy used for configuring simulated social media spaces.}
\label{tab:feature-taxonomy-starred}
\begin{tabular}{>{\raggedright\arraybackslash}p{4.2cm} >{\raggedright\arraybackslash}p{4.5cm} >{\raggedright\arraybackslash}p{6.3cm}}
\toprule
\textbf{Feature Category} & \textbf{Type} & \textbf{Description} \\
\midrule

\multirow{2}{=}{\textbf{Commenting Structure} \\ \small(Specifies the visual and conversational structure of replies)} 
  & Nested Threads & Comments and replies are grouped in a hierarchical tree format. \\
  & Flat Threads & All comments appear in a single-level linear sequence. \\
\addlinespace

\multirow{3}{=}{\textbf{Reactions} \\ \small(Indicates feedback mechanisms for user content)} 
  & Like-only & Single, positive feedback option such as a thumbs-up. \\
  & Upvote/Downvote & Binary control over content visibility based on user consensus. \\
  & Expanded Reactions & Emoji-based emotional responses offering varied feedback types. \\
\addlinespace

\multirow{1}{=}{\textbf{Content Management} \\ \small(Describes user control over posted content)} 
  & Edit/Delete & Users can revise or remove their content after posting. \\
\addlinespace

\multirow{3}{=}{\textbf{Identity and Account Type} \\ \small(Defines how users present themselves)} 
  & Real-name & Identity is tied to the user’s real name (e.g., LinkedIn). \\
  & Pseudonymous & Users adopt a username or handle (e.g., Reddit). \\
  & Anonymous & No persistent identity is used (e.g., 4chan). \\
\addlinespace

\multirow{2}{=}{\textbf{Messaging Modes} \\ \small(Captures communication modality)} 
  & Private (1:1) & One-on-one direct messaging between users. \\
  & Group Messaging & Small group or multi-user conversational spaces. \\
\addlinespace

\multirow{2}{=}{\textbf{Audience Scope} \\ \small(Specifies who can see or respond to content)} 
  & Everyone & Content is publicly accessible to all users. \\
  & With Connection & Content is limited to accepted users such as friends or members. \\

\bottomrule
\end{tabular}
\end{table*}

\renewcommand{\arraystretch}{1.4}
\begin{table*}[h]
\centering
\caption{Advanced features and customization (LV3) feature taxonomy used for configuring simulated social media spaces.}
\label{tab:system-level-features-starred}
\begin{tabular}{>{\raggedright\arraybackslash}p{4.3cm} >{\raggedright\arraybackslash}p{4.5cm} >{\raggedright\arraybackslash}p{6.2cm}}
\toprule
\textbf{Feature Category} & \textbf{Type} & \textbf{Description} \\
\midrule

\multirow{2}{=}{\textbf{Ephemeral Content} \\ \small(Controls the permanence of content)} 
  & Yes & Content disappears automatically after a set time (e.g., Stories, Snaps). \\
  & No & Content remains until manually removed by the user. \\
\addlinespace

\multirow{2}{=}{\textbf{Content Visibility Control} \\ \small(Manages who can view a user’s content)} 
  & Public & Posts are visible to anyone on the platform. \\
  & Private & Visibility is limited to approved users or group members. \\
\addlinespace

\multirow{2}{=}{\textbf{Content Discovery} \\ \small(Determines recommendation logic)} 
  & Topic-based Suggestion & Suggested content based on selected interest areas (e.g., Reddit). \\
  & Popularity-based Suggestion & Recommendations based on engagement metrics like views or likes (e.g., TikTok). \\
\addlinespace

\multirow{2}{=}{\textbf{Networking and Privacy Controls} \\ \small(Tools for setting social boundaries)} 
  & Block / Mute & Restrict interaction or hide content from selected users. \\
  & Invite-only Access & Access restricted to users with explicit invitation (e.g., Slack). \\

\bottomrule
\end{tabular}
\end{table*}

\newpage
\section{Supplemental Materials: Instantiating Social Behaviors}

In this section, we provide additional specific details about the process for simulating each of the social behaviors used in the simulated social platforms. Behaviors were grouped into three categories: Activity, Engagement, and Updates. We treat each in turn.

\paragraph{Activity: Content Creation and Direct Interaction}
This category includes actions such as posting, commenting, and reacting-core behaviors that drive social content generation and interaction.

\subparagraph{\textbf{\textit{Post Generation.}}}  
When an agent creates a post, the content is generated using multiple layers of metaphor-informed context~--~\textit{Content Orientation}, \textit{Actor Type}, and  \textit{Communication Flow} attribute that influences the tone and dynamics of communication.
% First, the system reflects the \textit{Content Orientation} attribute by aligning the post topics with themes extracted from the user's metaphor. Next, the \textit{Actor Type} attribute guides how the social perspective or role of the agent is embedded into the post's narrative. The \textit{Communication Flow} attribute influences the tone and structure of communication--e.g., formal vs. casual, thoughtful vs. emotional--selected from a predefined list of tones (~\ref{Appendix}) to match the intended interaction dynamics.

To ensure content diversity and avoid repetition, we impose both lexical and semantic variation constraints. Specifically, newly generated posts must share less than 20\% lexical overlap and have a cosine similarity below 0.2 compared to the agent's three most recent posts. This promotes varied and contextually rich content generation.

For channel-based posts, additional context of channel information including the name and bio are incorporated into the prompt. Channels are selected randomly among available groups. For ephemeral posts, the generated content is intentionally designed to reflect the informal and transient nature of this format. These posts are prompted to be shorter, more emotionally expressive, spontaneous in tone, and lack polished structure.

\subparagraph{\textbf{\textit{Commenting.}}}  
Comment generation considers the agent's relationship with the target post author and the comment writer. A closeness value is embedded into the content generation process, influencing the level of familiarity or intimacy in the response. Additionally, to maintain linguistic diversity, comments are constrained to have less than 30\% lexical overlap with the previous three comments. The content of the reply also takes into account the semantic context of the post or comment being responded to.

\subparagraph{\textbf{\textit{Reacting.}}}  
To simulate reaction behaviors, the system first selects a post at random from the agent's feed or visible content pool. Based on the platform's supported reaction types--such as likes, upvotes/downvotes, or emoji-based responses--the agent selects a reaction that fits the emotional or thematic content of the post. The choice of reaction also reflects the agent's role.

\paragraph{Engagement: Messaging and communication engagement}
This category of behaviors captures how agents engage in interpersonal and group communication, including one-on-one messaging and participation in community-driven channels. These actions are informed by metaphor-derived attributes such as audience type, relational closeness, tone, and communication flow.

\subparagraph{\textbf{\textit{Direct Messages.}}}  
To initiate a direct message, an agent randomly selects a target user from its network or from the entire user pool, depending on the messaging audience type inferred from the metaphor. Group chats follow a similar process but include multiple participants, with closeness levels embedded to reflect the intended social dynamic (e.g., intimate vs. casual). Once chats are created, agents send messages by randomly selecting a conversation and contributing only if the last message was not already theirs, thereby avoiding unrealistic back-to-back posting. To simulate natural conversation dynamics, approximately 10\% of messages include a short off-topic comment. Agents also read messages from their unread chat list based on a probabilistic trigger, supporting notification updates.

\subparagraph{\textbf{\textit{Channels.}}}  
When creating new community channels, agents are guided by metaphor-derived attributes to generate contextually appropriate channel details. For example, the channel's thematic focus is determined by the \textit{ContentOrientation} attribute, and the participants reflect the \textit{ActorType}. 
% A channel bio is generated to match the tone described by the \textit{Atmosphere} attribute, while the internal communication style follows the \textit{CommunicationFlow} attribute. 
To avoid redundancy in community creation, we apply a similarity threshold of 0.7 using the Jaro-Winkler distance metric; new channel names exceeding this threshold in similarity to existing ones are filtered out to promote topical diversity within the simulated space. When joining a channel, agents select from available communities based on their assigned personality type, ensuring that participation aligns with the social identity and interests inferred from the metaphor.

\paragraph{Updates: Changes to relationships and visibility}
To simulate evolving social dynamics and privacy preferences, agents periodically modify their relationships and visibility settings.

For relationship updates, agents randomly send or accept friend requests from a list of recommended users. These recommendations are generated based on overlapping topical interests, reflecting plausible social affinity. Once connections are established, agents may also update the status of existing relationships.

To simulate social boundaries and personal curation, agents may block or mute other users in contexts where network control is supported by the simulated environment. Additionally, agents adjust the visibility settings of their own posts. At random intervals, they may change content access from public to private (or vice versa), enabling a more dynamic simulation of privacy management behaviors common in social media environments.

\section{Supplemental Materials: Prompts for the Simulation Pipeline}
\label{sec:prompt-library}

This section documents the textual prompts used across the simulation pipeline. Placeholders (e.g., \verb|${user_id}|) are populated at runtime. We group prompts by function for clarity.

\subsection*{Metaphor Conversion}
\label{subsec:metaphor-conversion}

\begin{lstlisting}[caption={Converting metaphor into key attributes}, label={lst:metaphor-conversion-1}]
Given the metaphor keyword "${metaphorKeyword}", analyze it based on these attributes and return ONLY a JSON object with the following structure:
{
    "Atmosphere": "...",
    "GatheringType": "...",
    "ConnectingEnvironment": "...",
    "TemporalEngagement": "...",
    "CommunicationFlow": "...",
    "ActorType": "...",
    "ContentOrientation": "...",
    "ParticipationControl": "..."
  }

  Consider these definitions when analyzing:
  - Atmosphere: emotional and sensory qualities of the space
  - GatheringType: reason people come together (thematic or relation-based)
  - ConnectingEnvironment: how the space facilitates connections
  - TemporalEngagement: duration and frequency of participation
  - CommunicationFlow: interaction style and patterns
  - ActorType: type of social identity individuals adopt
  - ContentOrientation: dominant focus of communication
  - ParticipationControl: extent of visibility and interaction management

  Return ONLY the JSON object, no additional text or explanation.
\end{lstlisting}

\noindent

\begin{lstlisting}[caption={Mapping key attributes into social media features}, label={lst:metaphor-conversion-2}]
Based on these attributes: ${JSON.stringify(attributes)}, 
provide social media features organized in the following format:

LV1: Network Structure
- **Timeline Types**: Define how content is organized for users.
- Feed-based: Aggregates posts into a single scrolling interface (e.g., Facebook, Instagram).
- Chat-based: Segments conversations into thematic spaces or threads by using messages instead of posts (e.g., Slack, Discord).
- **Content Order**: Specifies the arrangement of content users see.
- Chronological: Content is displayed in the order it is posted (e.g., Twitter's "Latest Tweets" view).
- Algorithmic: Content is displayed based on relevance or popularity (e.g., Instagram, TikTok).
- **Connection Type**: Defines how users are connected and interact.
- Network-based: Connections between individuals such as friends or followers (e.g., Instagram, Twitter).
- Group-based: Collective participation within a predefined community (e.g., Reddit, Slack channels).
- **User Count**: Defines the exact number of users on the platform. Don't just pick middle number. think about attributes and the number of users it should have. The number should be minimum 5, and maximum 100. 

LV2: Interaction Mechanisms
- **Commenting**: Determines how users can respond to content.
- Flat Threads: Comments are displayed as a single-layered list.
- Nested Threads: Replies to comments are structured in a hierarchy.
- **Reactions**: Enables users to express their opinion on content.
- Like: A single positive acknowledgment (e.g., heart on Instagram posts).
- Upvote/Downvote: Allows for ranking content positively or negatively (e.g., Reddit).
- Expanded Reactions: Use of emojis such as "love," "haha," "angry," etc. (e.g., Facebook's reaction system).
- **Content Management**: Outlines options for editing or removing posts.
- Edit: Modify content after posting (e.g., X/Twitter edit feature for subscribers).
- Delete: Permanently remove content from the platform.
- **Account Types**: Defines privacy and accessibility. (multiple can be selected)
- Public: Content is accessible to everyone.
- Private (one-way): Follower requests are required, but users don't need mutual consent (e.g., Instagram private accounts).
- Private (mutual): Both parties must agree to connect (e.g., LinkedIn).
- **Identity Options**: Specifies how users represent themselves.
- Real-name: Users must use their real identity (e.g., LinkedIn).
- Pseudonymous: Users can use aliases (e.g., Instagram).
- Anonymous: Users are not identified (e.g., 4chan, Whisper).
- **Messaging**:
- Types: (multiple can be selected)
  - Private one-on-one (e.g., Facebook Messenger) 
  - group messaging (e.g., WhatsApp groups).
- Audience: You can message with people who have connection to you or everyone.
  - With connection 
  - everyone.

LV3: Advanced Features & Customization
- **Ephemeral Content**: Temporary content that disappears after a set time.
- Enabled: Platforms like Snapchat or Instagram Stories. (just reply with Yes or No)
- **Content Visibility Control**: Defines audience customization options. (choose between Public or Private)
- Public: Content is visible to all users
- Private: Content visibility is restricted
- **Content Discovery**: Methods of introducing users to new content.
- Recommendations:
- Topic-based Suggestions: Recommendations based on user interests (e.g., Pinterest).
- Popularity-based Suggestions: Recommendations based on trending content (e.g., TikTok's "For You" page).
- **Networking Control**: Tools to manage social interactions. (multiple can be selected)
- Block: Prevents another user from interacting with you
- Mute: Silences another user without notifying them
- **Privacy Settings**: Configures boundaries for interactions.
- Invited Content Only: Access is limited to invited users (e.g., Slack).
- Show All: Content is publicly visible to anyone (e.g., Instagram).

The answer structure should look like something like this:

LV1: Network Structure
Timeline Types: Chat-based
Content Order: Algorithmic
and so on...

Then at the end of the response, can you add your reasoning for the answer? Give specific reasoning for all your selections.

Do not use bolded text or []
\end{lstlisting}

\subsection*{Chat Generation}
\label{subsec:chat-generation}

\begin{lstlisting}[caption={Chat generation (dyadic).},label={lst:chat-dyadic}]
There is an ongoing conversation between two people. The last messages were:
"${formattedMessages}"

Context:
- Your user_id is ${user_id}.
- There is 1 other person in the chat.
- Your closeness level to the other person (1-10) is: ${closeness_levels}.

Goals:
- Respond naturally and personally to the last message.
- Do not repeat phrases or sentiments from earlier messages.
- You can use common chat shortforms or slangs like wby, love, luv, ngl, lol, lmao
- Try to keep the conversation engaging and personal. You may ask a follow-up question, express your opinion, or share a new idea.
- Limit your response to 1-2 short sentences, with no more than 12 words per message.
- Build on the conversation and ask deeper questions on the topic being discussed. Ensure the conversation flows naturally and builds upon the core topic in the last messages. For example, if someone is talking about food, give an example of a specific food you just ate. If someone asks "what's up?", reply with what you did that day (e.g., attended a class on business studies).
- About 10% of the time, include a one-line off-topic quip (a meme, weekend plan, news headline, etc.) unrelated to the main thread.
- If there is any question in the chat, reply to it before asking more questions.

Now, generate the next message as a single bubble.
\end{lstlisting}

\noindent

\begin{lstlisting}[caption={Chat generation (group).},label={lst:chat-group}]
There is an ongoing group chat. The last messages were:
"${formattedMessages}"

Context:
- Your user_id is ${user_id}.
- There are ${people.length} other people in the chat.
- Your closeness levels to them (1-10) are: ${closeness_levels}.

Goals:
- Respond naturally, but keep in mind this is a group conversation. You may reference others, introduce new topics, or ask general questions.
- Do not repeat phrases or sentiments from earlier messages.
- Keep the conversation varied. Introduce new angles, switch the tone, or share a new topic.
- Limit your response to 1-2 short sentences, with no more than 12 words per message.
- Avoid using an exclamation mark unless absolutely necessary.
- Build on the conversation and ask deeper questions on the topic being discussed. Ensure the conversation flows naturally and builds upon the core topic in the last messages. For example, if someone is talking about food, give an example of a specific food you just ate. If someone asks "what's up?", reply with what you did that day (e.g., attended a class on business studies).
- About 10% of the time, include a one-line off-topic quip (a meme, weekend plan, news headline, etc.) unrelated to the main thread.
- If there is any question in the chat, reply to it before asking more questions.

Now, generate the next message(s) as separate bubbles.
\end{lstlisting}

\subsection*{Post Generation}
\label{subsec:post-generation}

\begin{lstlisting}[caption={Personal post generation (non-ephemeral).},label={lst:post-personal}]
You are a user on social media platforms like ${platforms.join(", ")}.
When writing a new post, mimic the typical style of that platform in terms of:
- Length (120-150 characters, max three sentences) and tone (avoid exclamation marks unless necessary)
- Formatting (informal, no bullet points, no bold/italic, use natural paragraph breaks)
- Hashtag use (use minimal, aligning to the platform's culture, don't overdo it)
Do NOT sound like a corporate announcement or a generic AI.

POST CONTENT REQUIREMENTS:
0. The post content MUST reflect topics related to ${descr.llm_descr.ContentOrientation} that may arise from interactions among ${descr.llm_descr.ActorType}.

1. Select a tone from the list (${tone}) that best matches the style of ${descr.llm_descr.CommunicationFlow}
2. Pick one user goal from ${user_roles} and generate a post based on the behavior associated with that goal.
3. Your post must be significantly different from your last three posts in:
   - Content, structure, storyline, length, and phrasing
   - Lexical overlap: below 20% shared words with past 3 posts
   - Semantic similarity: below 0.2 cosine similarity with past 3 posts. Use a completely different sentence structure.
   The contents of some of your previous posts are: ${last_posts}.
4. Structure the post clearly with natural newlines-avoid dense blocks of text.
5. Keep the contents engaging and relatable.
6. Avoid generic tone if your last two posts were already generic-add specificity (names, places, small moments).
7. Do not end the post with a question.
8. Do NOT start the sentence with words like "JUST", "FINALLY", "FOUND", "HAD", "CURRENTLY", "CAME ACORSS".

Now, generate a new post that sticks to a single theme and meets all of the above criteria.
\end{lstlisting}

\noindent

\begin{lstlisting}[caption={Personal post generation (ephemeral).},label={lst:post-personal-ephemeral}]
You are a user on social media platforms like ${platforms.join(", ")}.
You are about to make a new **ephemeral** post on social media. These are time-sensitive posts and will only be up for 24 hours.
When writing a new ephemeral post, mimic the typical style of that platform in terms of:
- Short and concise length (30~40 characters, max two sentences) 
- Informal, spontaneous, or unpolished tone (avoid exclamation marks unless necessary)
- Personal and emotionally expressive
- Formatting (no bullet points, no bold/italic, use natural paragraph breaks)
Do NOT sound like a corporate announcement or a generic AI. 

POST CONTENT REQUIREMENTS: 
0. The post content MUST reflect topics related to content orientation in ${descr.lvl1.llm_descr} that may arise from interactions among actor type in ${descr.lvl1.llm_descr}.           
1. Select a tone from the list (${tone}) that best matches the style of ${descr.llm_descr.CommunicationFlow}
2. "Pick one user goal from ${user_roles} and generate a post based on the behavior associated with that goal.
3. Your post must be significantly different from your last three posts in:
- Content, structure, storyline, length, and phrasing
- Lexical overlap: below 20% shared words with past 3 posts
- Semantic similarity: below 0.2 cosine similarity with past 3 posts. Use a completely different sentence structure. The contents of some of your previous posts are:${last_posts}. 
4. Structure the post clearly with natural newlines-avoid dense blocks of text.
5. Keep the contents engaging and relatable.
6. Avoid generic tone if your last two posts were already generic-add specificity (names, places, small moments).
7. Do not end the post with a question.
8. Do NOT start the sentence with words like "JUST", "FINALLY", "FOUND", "HAD", "CURRENTLY", "CAME ACORSS".

Now, generate a new post that sticks to a single theme and meets all of the above criteria.
\end{lstlisting}

\subsection*{Channel Post Generation (Community))}
\label{subsec:channel-post-generation}

\begin{lstlisting}[caption={Community-aligned channel post generation (non-ephemeral).},label={lst:post-channel}]
You are about to make a new post in a community.
The community name is ${sel_comm.comm_name}. This is a community with likeminded people who are passionate about ${sel_comm.comm_bio}.
You are a user on social media platforms like ${platforms.join(', ')}.
When writing a new post, mimic the typical style of that platform in terms of:
- Length (120-150 characters, max three sentences) and tone (avoid exclamation marks unless necessary)
- Formatting (informal, no bullet points, no bold/italic, use natural paragraph breaks)
- Hashtag use (use minimal, aligning to the platform's culture, don't overdo it)
Do NOT sound like a corporate announcement or a generic AI.

POST CONTENT REQUIREMENTS:
0. The post content MUST reflect topics related to ${descr.llm_descr.ContentOrientation} that may arise from interactions among ${descr.llm_descr.ActorType}.

1. Your post must be aligned with the community topic.
2. Select a tone from the list (${tone}) that best matches the style of ${descr.llm_descr.CommunicationFlow}
3. Pick one theme among the user interests: ${user_interests}. Focus on one clear theme. Do not mix unrelated ideas.
4. Pick one user goal from ${user_roles} and generate a post based on the behavior associated with that goal.
5. Your post must be significantly different from your last three posts in:
   - Content, structure, storyline, length, and phrasing
   - Lexical overlap: below 20% shared words with past 3 posts
   - Semantic similarity: below 0.2 cosine similarity with past 3 posts. Use a completely different sentence structure.
   The contents of some of your previous posts are: ${last_posts}.
6. Structure the post clearly with natural newlines-avoid dense blocks of text.
7. Keep the contents engaging and relatable.
8. Avoid generic tone if your last two posts were already generic-add specificity (names, places, small moments).
9. Do not end the post with a question.
10. Do NOT start the sentence with words like "JUST", "FINALLY", "FOUND", "HAD", "CURRENTLY", "CAME ACORSS".

Now, generate a new post that sticks to a single theme and meets all of the above criteria.
\end{lstlisting}

\noindent

\begin{lstlisting}[caption={Community-aligned channel post generation (ephemeral).},label={lst:post-channel-ephemeral}]
You are about to make a new **ephemeral post** on social media. These are time-sensitive posts and will only be up for 24 hours.
The community name is ${sel_comm.comm_name}. This is a community with likeminded people who are passionate about ${sel_comm.comm_bio}.

You are a user on social media platforms like ${platforms.join(", ")}.
When writing a new ephemeral post, mimic the typical style of that platform in terms of:
- Short and concise length (30~40 characters, max two sentences) 
- Informal, spontaneous, or unpolished tone (avoid exclamation marks unless necessary)
- Personal and emotionally expressive
- Formatting (no bullet points, no bold/italic, use natural paragraph breaks)
Do NOT sound like a corporate announcement or a generic AI. 

POST CONTENT REQUIREMENTS: 
0. The post content MUST reflect topics related to ${descr.llm_descr.ContentOrientation} that may arise from interactions among ${descr.llm_descr.ActorType}.           
1. Your post must be aligned with the community topic.
2. Select a tone from the list (${tone}) that best matches the style of ${descr.llm_descr.CommunicationFlow}
3. Pick one theme among the user iterests: ${user_interests}. Focus on one clear theme. Do not mix unrelated ideas.
4. "Pick one user goal from ${user_roles} and generate a post based on the behavior associated with that goal.
5. Your post must be significantly different from your last three posts in:
- Content, structure, storyline, length, and phrasing
- Lexical overlap: below 20% shared words with past 3 posts
- Semantic similarity: below 0.2 cosine similarity with past 3 posts. Use a completely different sentence structure. The contents of some of your previous posts are:${last_posts}. 
6. Structure the post clearly with natural newlines-avoid dense blocks of text.
7. Keep the contents engaging and relatable.
8. Avoid generic tone if your last two posts were already generic-add specificity (names, places, small moments).
9. Do not end the post with a question.
10. Do NOT start the sentence with words like "JUST", "FINALLY", "FOUND", "HAD", "CURRENTLY", "CAME ACORSS".

Now, generate a new post that sticks to a single theme and meets all of the above criteria.
    
\end{lstlisting}

\subsection*{Community Selection (Join Channel)}
\label{subsec:join-channel}

\begin{lstlisting}[caption={Join a community (selection-only response).},label={lst:join-channel}]
You want to join a new community. Based on your personality, choose ONE from the list below.
Be direct and reply with ONLY the community ID of the selected community.

Available communities:
[... runtime-provided list with IDs ...]
\end{lstlisting}

\subsection*{Agent Generation from Metaphor}
\label{subsec:agent-from-metaphor}

\paragraph{System message.}
\noindent
\begin{lstlisting}[caption={Agent generation (system).},label={lst:agent-system}]
You are an AI that generates social media user profiles based on metaphorical descriptions.
The user has the goal of "${goalRole.goal}" and plays the role of "${goalRole.role}".
Create a personality that embodies these metaphorical characteristics:
LLM Description: ${llm_descr}
\end{lstlisting}

\paragraph{User message (JSON-spec output).}
\noindent
\begin{lstlisting}[caption={Agent generation (user with JSON schema).},label={lst:agent-user}]
Create a social media user profile that embodies the goal of "${goalRole.goal}" and the role of "${goalRole.role}".

USERNAME REQUIREMENTS:
- Strictly follow this identity style: ${identity_prompt}
- CRUCIAL: If the identity type is psedononymous, the username MUST be somehow related to the metaphorical theme '${descriptions.keyword}. It doesn't need to include the metaphor keyword itself.'.
- Please follow the universal and standard naming convention used in general social media.
${existingUserNames.length > 0 ? `
- ABSOLUTELY ESSENTIAL: The username MUST be different from these existing names:
  ${existingUserNames.join(', ')}` : ''}

Generate a JSON object with these required fields:
{
  "id_name": "A unique identifier starting with 'ID_'",
  "user_name": "A username strictly adhering to the USERNAME REQUIREMENTS above.",
  "email": "A thematic email address, can be related to the metaphor or username strategy",
  "password": "A strong password",
  "user_bio": "A concise (1-3 sentences, approx. 150 characters), engaging social media bio that reflects the general writing style of typical social media bios. This bio should relate to ${llm_descr.ContentOrientation} content and reflect the vibe of ${llm_descr.Atmosphere}, where users are gathered around the ${llm_descr.GatheringType} theme. This bio should NOT weave in the metaphorical theme of '${descriptions.keyword} or metaphor.' ${existingUserBios.length > 0 ? `It MUST be distinct from these existing bios: ${existingUserBios.map(bio => `"${bio}"`).join('; ')}. ` : ''}No emojis.",
  "profile_picture": "A URL using https://i.pravatar.cc/120?u= with a random parameter",
  "posting_trait": "Float between 0-1",
  "commenting_trait": "Float between 0.5-1",
  "reacting_trait": "Float between 0.5-1",
  "messaging_trait": "Float between 0.5-1",
  "updating_trait": "Float between 0-1",
  "comm_trait": "Float between 0-1",
  "notification_trait": "Float between 0-1",
  "interests": ["At least 3 interests from the predefined list that align with ${llm_descr.ContentOrientation} contents"],
  "persona_name": "Name the user's personality type that appears from ${llm_descr.ActorType} in ${llm_descr.ConnectingEnvironment} social connecting environment. Should NOT be making the real name.",
  "social_group_name": "A group name aligned with the metaphor. Make sure it ranges in tone, length and nuance."
}

Ensure the personality traits and interests align with the metaphorical description.
The predefined interests list: ["Animals", "Art & Design", "Automobiles", "DIY & Crafting", "Education", "Fashion", "Finance", "Fitness", "Food", "Gaming", "History & Culture", "Lifestyle", "Literature", "Movies", "Music", "Nature", "Personal Development", "Photography", "Psychology", "Religion", "Social", "Sports", "Technology", "Travel", "Wellness"]

Return only the JSON object.
\end{lstlisting}

\subsection*{Comment Generation}
\label{subsec:comment-generation}

\begin{lstlisting}[caption={Comment generation on a post.},label={lst:comment-generation}]
You are about to comment on a post. The content of the post is: "${sel_post.content}".
On a scale of 1 to 10, your closeness level with the person is "${closeness}".
Generate a comment for the post that is a one liner 60% of the time.
Leave an emoji only when it is absolutely necessary, not otherwise.
Vary your mood slightly: supportive, curious, witty, or reflective, but deliver it in a calm nonchalant way-don't be upbeat every time.
Dive deep into the post and talk about specific things related to the post.
Switch it up with small comments like "wow, good read" or "interesting perspective, I was thinking about this the other day".
Avoid using an exclamation mark unless absolutely necessary.
Ensure your phrasing is <30% lexically overlapping with any of your last 3 comments.
\end{lstlisting}

\label{subsec:placeholder-reference}
\noindent\textbf{Runtime placeholders.} 
\begin{multicols}{3}
\raggedright
\texttt{\${user\_id}} \\
\texttt{\${people.length}} \\
\texttt{\${closeness\_levels}} \\
\texttt{\${formattedMessages}} \\
\texttt{\${platforms}} \\
\texttt{\${tone}} \\
\texttt{\${user\_roles}} \\
\texttt{\${last\_posts}} \\
\texttt{\${sel\_comm.comm\_name}} \\
\texttt{\${sel\_comm.comm\_bio}} \\
\texttt{\${user\_interests}} \\
\texttt{\${descr.llm\_descr.*}} \\
\texttt{\${goalRole.*}} \\
\texttt{\${identity\_prompt}} \\
\texttt{\${existingUserNames}} \\
\texttt{\${existingUserBios}} \\
\texttt{\${descriptions.keyword}} \\
\texttt{\${sel\_post.content}} \\
\texttt{\${closeness}}
\end{multicols}

\end{document}